\documentclass[fleqn,usenatbib]{rasti}

\usepackage{newtxtext,newtxmath}
\usepackage[T1]{fontenc}

\DeclareRobustCommand{\VAN}[3]{#2}
\let\VANthebibliography\thebibliography
\def\thebibliography{\DeclareRobustCommand{\VAN}[3]{##3}\VANthebibliography}

\usepackage{graphicx}	% Including figure files
\usepackage{amsmath}	% Advanced maths commands
\usepackage{booktabs}
\usepackage{hyperref}
\usepackage{upgreek}

\title[Detecting Diffuse Cluster Emissions with Astronomaly]{A targeted machine learning approach for detecting diffuse radio emission with Astronomaly: Protege}

\author[Verlon Etsebeth et al.]{
Verlon Etsebeth$^{1}$\thanks{E-mail: verlon18@gmail.com},
Michelle Lochner$^{1}$, 
Konstantinos Kolokythas$^{2,3}$,
Kenda Knowles$^{2,3}$,
Emma Tolley$^{4}$
\vspace{1.5ex} \\ 
$^{1}$Department of Physics and Astronomy, University of the Western Cape, Bellville, Cape Town 7535, South Africa\\
$^{2}$Centre for Radio Astronomy Techniques and Technologies, Department of Physics and Electronics, Rhodes University, \\
P.O. Box 94, Makhanda 6140, South Africa\\
$^{3}$South African Radio Astronomy Observatory, Liesbeek House, River Park, Gloucester Road, Mowbray, Cape Town 7700, South Africa \\
$^{4}$Institute of Physics, Laboratory of Astrophysics, \'Ecole Polytechnique F\'ed\'erale de Lausanne (EPFL), Observatoire de Sauverny, Versoix 1290, Switzerland
}

\pubyear{\the\year{}}

\begin{document}
\label{firstpage}
\pagerange{\pageref{firstpage}--\pageref{lastpage}}
\maketitle

\begin{abstract}
    Diffuse radio emission in galaxy clusters, such as radio halos, relics, and mini halos, is a key tracer of non-thermal processes, turbulence, and magnetic fields within the intra-cluster medium. However, their low surface brightness, as well as contamination from compact sources and imaging artefacts, makes their detection challenging. The sheer volume of data from instruments such as the Square Kilometre Array will render traditional manual-inspection based detection methods infeasible. This paper introduces a novel machine learning approach that uses active learning to rapidly identify diffuse emission candidates from a small, optimally-selected subset of data. We apply the self-supervised deep learning algorithm Bootstrap Your Own Latent to extract features from source cutouts in the MeerKAT Galaxy Cluster Legacy Survey (MGCLS). We then pass these features through the \textsc{Astronomaly: Protege} anomaly detection framework to identify the final candidates. Using a human-labelled set, we evaluate our pipeline on high-resolution ($\sim7\arcsec$), convolved (15\arcsec), and combined-feature MGCLS datasets. Interestingly, the high-resolution features identify diffuse sources more efficiently than the convolved resolution, which are in turn outperformed by the combined features. Of the top 100 sources ranked by \textsc{Protege}, 98\% exhibit diffuse characteristics, with 53\% confirmed as cluster-related emission. Our work shows that \textsc{Protege} can identify diffuse emission with minimal human labelling effort, offering a powerful, scalable tool capable of detecting both known and novel diffuse radio sources.
\end{abstract}

% Include between one and six keywords.

%%%%%%%%%%%%%%%%% BODY OF PAPER %%%%%%%%%%%%%%%%%%

%
%%
%%% ====== Section
\section{Introduction}
\label{sec:introduction}
%%% ===============
%%
%

Galaxy clusters, as the largest gravitationally bound structures in the Universe, offer a valuable means of studying cosmic evolution. The intra-cluster medium (ICM) within these systems hosts a variety of thermal and non-thermal phenomena, shaped by the interplay between turbulence, magnetic fields, and particle acceleration processes \citep{brunetti2014nonthermal,vanWeeren2019}. Diffuse radio emission found within galaxy clusters originates from synchrotron radiation produced by relativistic electrons interacting with magnetic fields and trace the dynamic, non-thermal components of the ICM. They can therefore constrain theories of cosmic-ray transport and magnetic field evolution \citep{Knowles2021}. Detecting new diffuse emission structures across diverse cluster environments is therefore important for understanding the origin and physical drivers of these phenomena.

Diffuse radio \textit{cluster} emission falls into three main classes based on morphology, location, and origin \citep{brunetti2014nonthermal, vanWeeren2019, 2021A&A...651A.115V, 2022A&A...665A..60H, 2022A&A...660A..78B, kolokythas2025}. \textit{Radio halos} appear as megaparsec-scale emission coinciding with the centres of galaxy clusters, overlapping with the X-ray emitting ICM, and are linked to merger-driven turbulence. \textit{Radio relics} manifest as large, highly polarised structures at cluster peripheries and trace shock fronts from cluster mergers. \textit{Radio mini-halos}, typically $\le500$ kpc, occupy central regions in cool-core, dynamically relaxed clusters and arise from gas sloshing and particle re-acceleration \citep{Giacintucci_2017}. Recent observations have also revealed complex structures such as radio phoenixes (revived fossil plasma) and faint emissions bridging clusters, which further contribute to understanding large-scale cosmic web formation \citep{Cuciti2021a, Venturi_radio_bridge}. However, diffuse emission can also occur outside of clusters from sources such as dying radio galaxies or remnant active galactic nuclei (AGN) \citep{2021A&A...648A..11O}.

Modern radio facilities like LOFAR, uGMRT, VLA and ASKAP have improved the detection of faint diffuse emission by offering high sensitivity and angular resolution across a range of redshifts and cluster masses \citep{Bonafede2011, vanWeeren2019, Cuciti2021a, 2021PASA...38....9H}. The MeerKAT Galaxy Cluster Legacy Survey (MGCLS), in particular, provided deep 1.28GHz observations with high sensitivity and resolution, ideal for studying these faint, extended structures \citep{Knowles_2022}. 

However, detecting these structures remains challenging due to their intrinsically low surface brightness and corresponding low signal-to-noise ratios (SNRs). Even with modern instruments, identifying and classifying diffuse cluster emission is complicated by contamination from compact sources such as AGN, source blending, and imaging artefacts \citep{kolokythas2025}. Additionally, the sheer volume and complexity of data anticipated from next-generation instruments such as the SKA will render manual inspection impractical.

To address these challenges, machine learning offers scalable automation. Several approaches have emerged recently: Transformer-based segmentation models like \textsc{Tuna}, which uses a customised TransUNet architecture \citep{chen2021transunettransformersmakestrong}, employs synthetic training data to map faint extended sources in LoTSS \citep{Sanvitale_2025}, while multimodal foundation models such as OpenCLIP enable similarity-based searches for peculiar radio galaxy morphologies in the Evolutionary Map of the Universe (EMU) survey \citep{gupta2025emuseevolutionarymapuniverse}. However, these methods require either large labelled datasets or synthetic simulations matched to the target survey, resources that may not exist for exploratory surveys where target morphologies are poorly constrained or can not be defined in advance.

Other approaches include anomaly detection, but this often identifies uninteresting outliers while burying scientifically valuable sources in complex feature spaces \citep{lochner2024}, as demonstrated by sources like SAURON being overlooked in MeerKAT data \citep{Lochner_2023}. The \textsc{Astronomaly} framework addresses this limitation by combining unsupervised anomaly detection with active learning, thereby personalising results by learning which anomalies match user preferences rather than simply flagging all outliers \citep{Lochner2021}.

Active learning, as implemented in \textsc{Astronomaly} and its extension \textsc{Protege} \citep{lochner2024}, operates differently than standard supervised machine learning. Rather than training a model on one dataset to apply to another, the algorithm works iteratively within a single dataset to identify sources of interest. There is no separate training and test split because the goal is not generalisation but rather efficient exploration of the specific dataset at hand. The algorithm continuously learns from human labels and uses this feedback to rank sources, with performance measured by how effectively interesting sources rise to the top of the ranked list. This makes \textsc{Protege} particularly well-suited for scenarios where the targets are rare, morphologically diverse, and can not be easily defined in advance.

In this work, we combine self-supervised feature extraction using Bootstrap Your Own Latent (BYOL, \citealp{grill2020bootstrap}) with \textsc{Protege} for targeted source detection. BYOL learns a compact, image-based feature representation of all sources without requiring labels, providing a structured feature space in which morphological similarities are encoded. Previous work has demonstrated that BYOL can automatically cluster galaxies with similar morphologies and highlight rare or unusual types, such as mergers \citep{Mohale2024}, illustrating its capability to capture meaningful features for scientific discovery. \textsc{Protege} then operates on this feature space as a recommendation engine, guiding the search based on user-defined criteria. This enables efficient discovery of diffuse emission candidates in MGCLS data without exhaustive manual inspection. Importantly, the approach emulates exploring a completely unlabelled dataset from scratch: the algorithm makes no assumptions about what constitutes diffuse emission or its prevalence, learning these distinctions entirely from user feedback.

We aim to (1) develop and test a pipeline for detecting diffuse cluster emission in a targeted manner, (2) compare the performance of high-resolution and convolved datasets, and (3) evaluate the benefits of concatenating feature representations from different resolutions.

We describe the MGCLS data in \autoref{sec:data}, detail source selection in \autoref{sec:source selection}, explain BYOL feature extraction and visualisation in \autoref{sec:feature extraction and visualisation}, and present the \textsc{Astronomaly: Protege} implementation in \autoref{sec:active learning with protege}. Pipeline performance is evaluated in \autoref{sec:results}, followed by discussion in \autoref{sec:discussion} and conclusions in \autoref{sec:conclusion}. A flowchart summarising the complete pipeline is provided in \autoref{sec:appendix flowchart}.

%
%%
%%% ====== Section
\section{Data}
\label{sec:data}
%%% ===============
%%
%

%
%% ====== Subsection
\subsection{The MeerKAT Galaxy Cluster Legacy Survey}
\label{sec:mgcls}
%% ===============
%

This work makes use of the MGCLS, one of the most detailed widefield radio surveys targeting galaxy clusters \citep{Knowles_2022}. MGCLS performed long-track L-band (900–1670 MHz) observations of 115 galaxy clusters, with each cluster observed for approximately 6 to 10 hours. The survey aims to study galaxy evolution, intra-cluster magnetism, cluster dynamics, star formation, and neutral hydrogen mapping in clusters.

The 64 antennas of MeerKAT, spanning 8-km with a densely packed 1-km core, provide high sensitivity images with typical RMS noise levels of $\sim3-5\upmu\mathrm{Jy} \text{beam}^{-1}$. MGCLS achieves angular resolutions ranging from approximately $\sim 7\arcsec-8\arcsec$ (high-resolution products) to $15\arcsec$ in convolved images, with the lower resolution considered better suited for detecting faint, extended emission. The survey is sensitive to structures up to $10{\arcmin}$ in angular scale.

The MGCLS also includes polarisation and spectral index image cubes, which can be used for Faraday rotation studies and mapping magnetic fields within clusters, though we do not use these products in our analysis. The basic image cubes cover a full field of view measuring $2^\circ\times2^\circ$, allowing for wide-field observations of galaxy clusters. In contrast, the enhanced products focus on the inner $1.2^\circ\times1.2^\circ$ field, which is corrected for the primary beam, and are used in this study.

%
%% ======  Subsection
\subsection{Human Labelled Catalogue}
\label{sec:human labelled catalogue}
%% ===============
%

\begin{figure*}
\centering
\begin{minipage}{\linewidth} 
    \centering
    \includegraphics[width=0.245\linewidth]{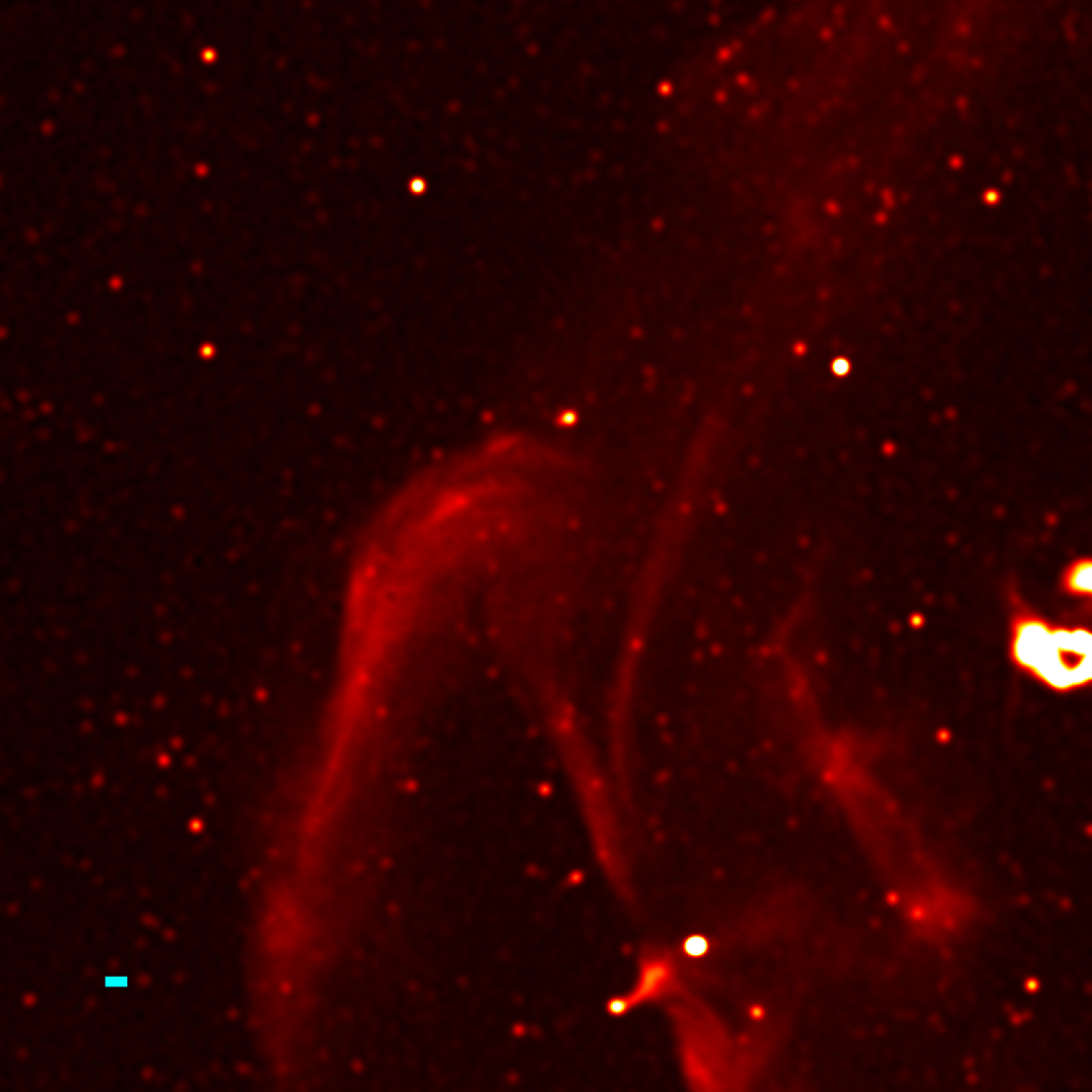}
    \includegraphics[width=0.245\linewidth]{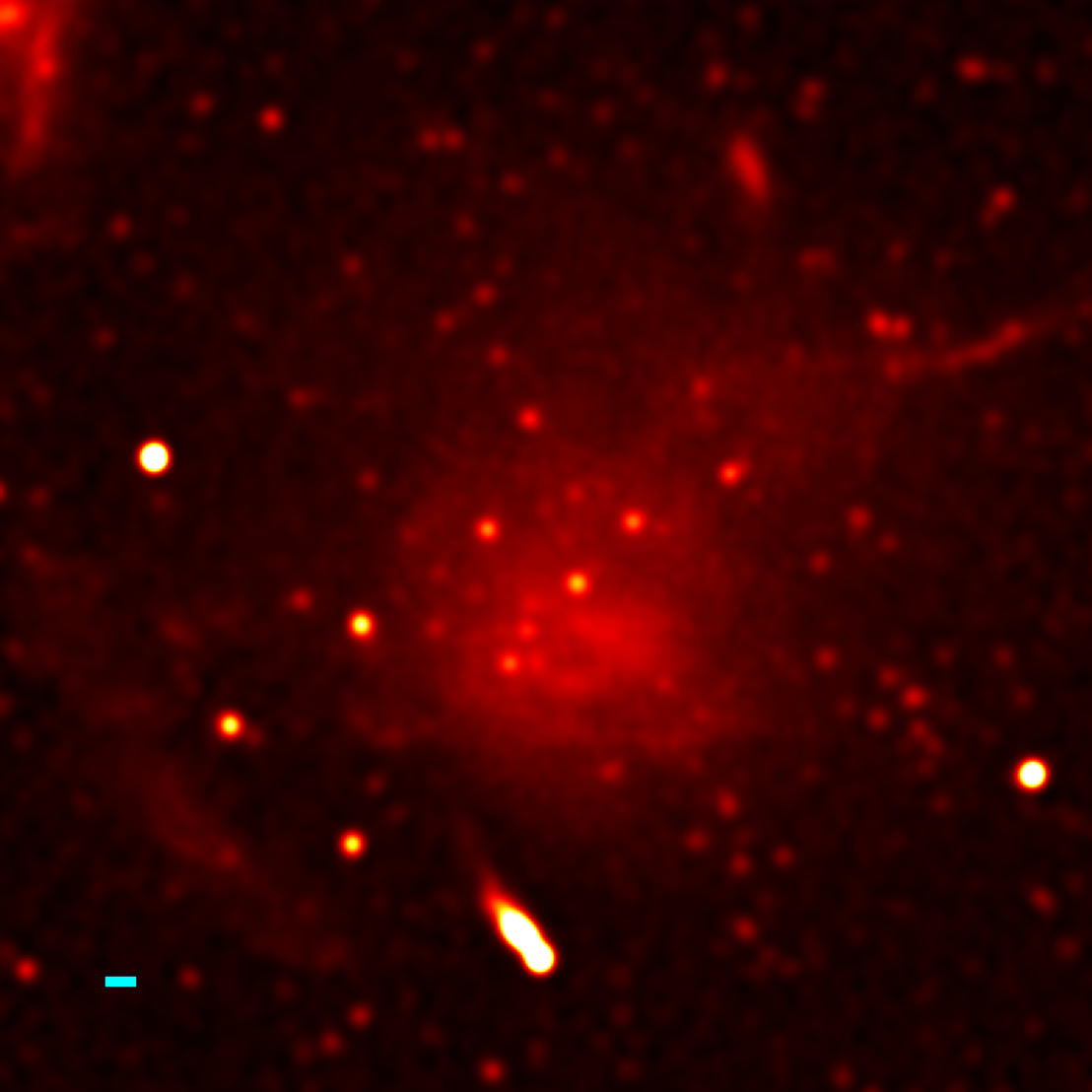}
    \includegraphics[width=0.245\linewidth]{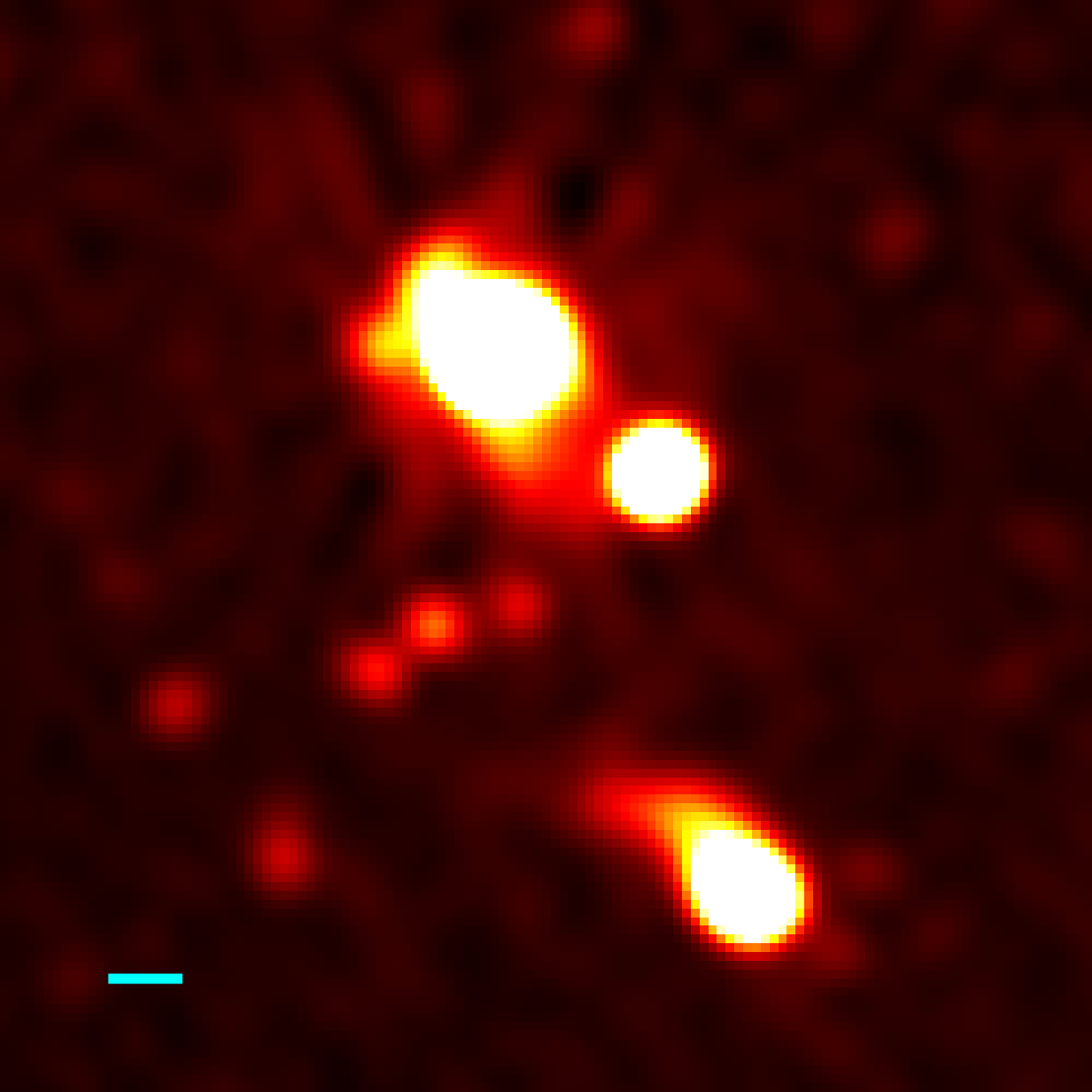}
    \includegraphics[width=0.245\linewidth]{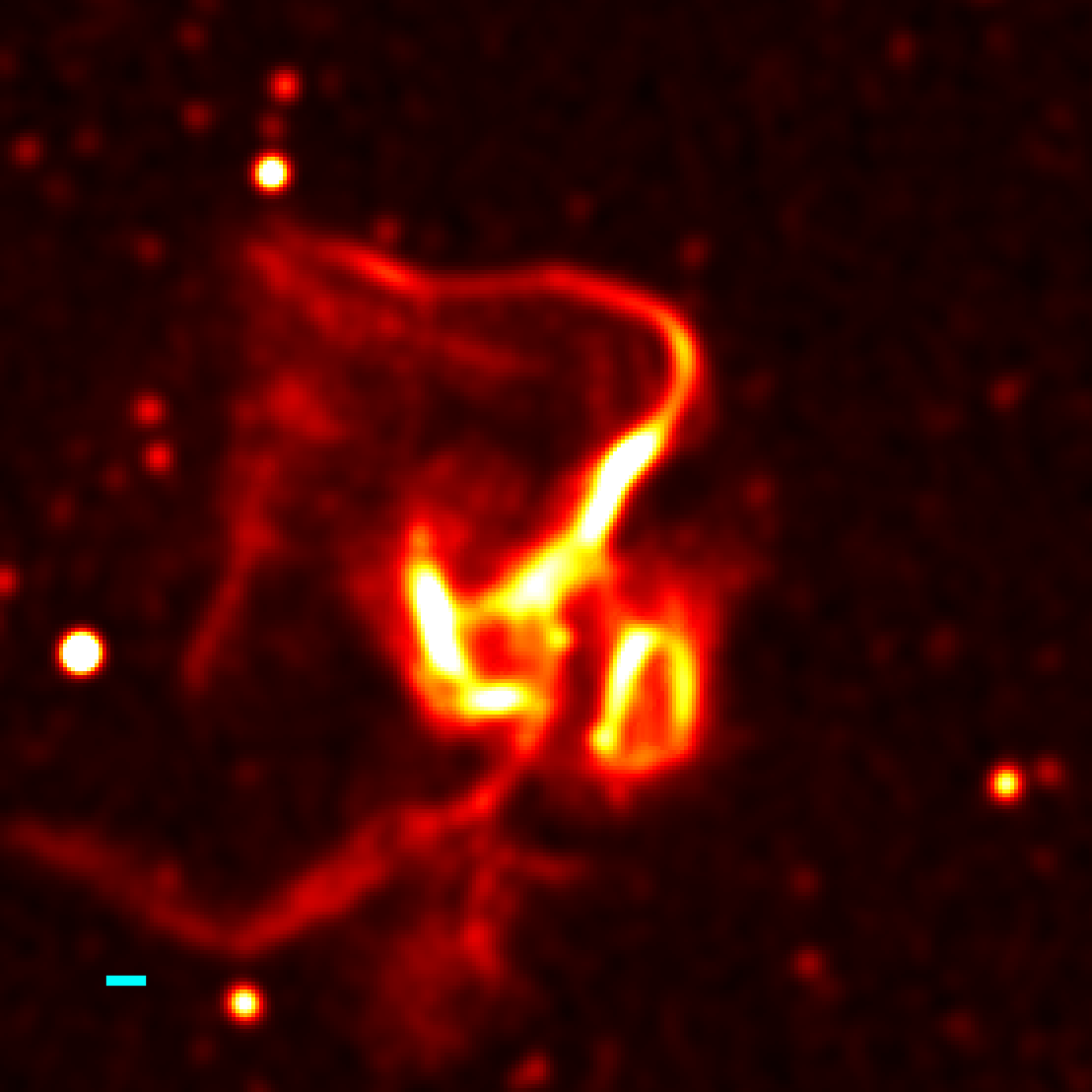}
    \caption{Examples of some of the known diffuse cluster emission in this study. From left to right: relic, halo, mini halo, and phoenix, ordered according to their frequency in the MGCLS catalogue. The beam solid angle is shown in the bottom-left corner of each cutout.}
    \label{fig:examples_of_tracers}
    Alt text: Four square radio cutouts coloured by a heatmap. The relic is a long, thin, curved arc running down through the entire cutout. The halo is a faint, round glow filling the centre of its image, with several small bright compact sources throughout the cutout. The mini halo is faint, patchy emission around a bright compact source. The phoenix is a bright, tangled, filament-like structure. A short cyan bar in the lower-left corner of each image marks the beam size.
\end{minipage}
\end{figure*}

\cite{kolokythas2025} provides a detailed catalogue of all 115 MGCLS clusters, reporting that approximately 54\% (62 of the 115 clusters) display some form of diffuse radio cluster emission, with some clusters hosting multiple sources. This catalogue, hereafter referred to as the Human Labelled Catalogue (HLC), represents a follow-up to the original MGCLS overview \citep{Knowles_2022} and provides a detailed characterisation of diffuse structures across both high-resolution and convolved low-resolution images. The catalogue refined classifications and identified new structures, increasing the total count from 99 to 103 diffuse radio sources or candidates. Examples of catalogued sources are shown in \autoref{fig:examples_of_tracers}. This expert-labelled catalogue is used throughout this work as a benchmark to evaluate the cumulative recall performance of our pipeline and to guide labelling during the active learning phase.

%------------------------------------------------%
%------------------------------------------------%

%
%% ====== Subsection
\subsection{Source Extraction}
\label{sec:source extraction}
%% ===============
%

We performed source extraction on the MeerKAT enhanced products using the Python Blob Detection and Source Finder (PyBDSF; \citealp{2015ascl.soft02007M}), a widely used tool in radio interferometry surveys. PyBDSF identifies contiguous regions of emission, termed ``islands,'' and fits Gaussian components to model these sources, generating detailed catalogues and managing background noise estimation. 

Radio catalogues typically use the Gaussian components to represent sources. However, to ensure diffuse emission is captured, we instead use the islands themselves to create the image cutouts used in subsequent analysis. This approach preserves extended structures that might otherwise be inadequately represented by a small number of Gaussian components.

We inspected a range of PyBDSF parameter choices, assessing their effect through the trade-off between retention of known sources from the HLC and the total number of extracted sources. As no configuration tested improved HLC source recovery without substantially increasing the resulting catalogue size within the parameters explored, we adopted the standard PyBDSF thresholds of \texttt{thresh\_isl=3.0} and \texttt{thresh\_pix=5.0}.

%
%%
%%% ====== Section
\section{Source Selection}
\label{sec:source selection}
%%% ===============
%%
%

PyBDSF source extraction on the high-resolution and convolved datasets yielded 62,587 and 40,879 cutouts, respectively. Most of these sources are compact and could be filtered out to reduce computational costs. However, any filtering needs to preserve the known diffuse emission sources from the HLC (\autoref{sec:human labelled catalogue}), as these are the primary targets of this study.

%
%% ====== Subsection
\subsection{Tracers}
\label{sec:tracers}
%% ===============
%

%------------------------------------------------%
\begin{table}
    \centering
    \begin{tabular}{llll}
        \toprule
        \textbf{Resolution} & \textbf{Source} & RA (J2000) & Dec. (J2000) \\ \midrule
        High Res. & Abell 2811 & 00:42:08.8 & –28:32:08.8 \\
                  & RXC J0520.7–1328 & 05:21:02.2 & –13:35:26.5 \\
                  &                  & 05:21:09.8 & –13:29:07.7 \\
                  &                  & 05:20:49.5 & –13:31:57.0 \\
                  & RXC J2351.0–1954 & 23:50:41.3 & –19:56:27.1 \\
                  & J0027.3–5015     & 00:27:21.3 & –50:15:04.0 \\ \midrule
        Convolved & RXC J0520.7–1328 & 05:21:02.2 & –13:35:26.5 \\
        \bottomrule
    \end{tabular}
    \caption{Tracers not detected and extracted using the default parameters of PyBDSF.}
    \label{tab:sources_not_found}
\end{table}
% %------------------------------------------------%

\begin{figure*}
\begin{minipage}{\textwidth}
    \centering
        \includegraphics[width=\textwidth]{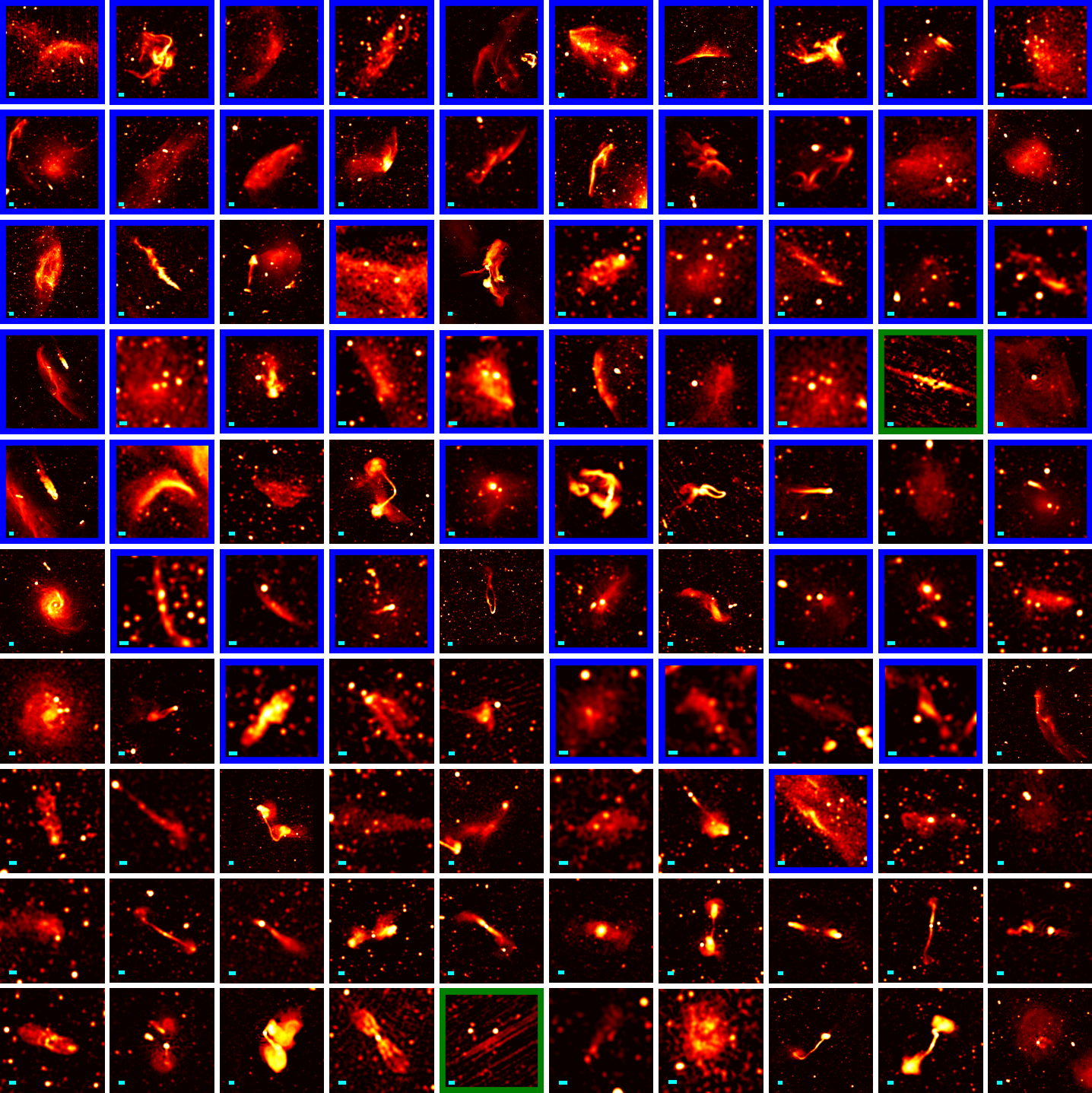}
    \caption{Top 100 ranked sources from the concatenated high-resolution subset (see \autoref{sec:concatenated features}). Each panel shows the 128$\times$128 pixel cutout size used as input to BYOL. Known tracers are highlighted with blue frames, while green frames indicate sources identified as imaging artefacts. The beam solid angle is shown in the bottom-left corner of each cutout. 98 of the 100 sources represent some form of diffuse and/or extended radio emission, highlighting the performance of the algorithm. A more detailed analysis of these sources is presented in \autoref{sec:top 100 investigation}.}
    \label{fig:top 100 high res concatenated}
    Alt text: A 10 by 10 grid of small cutouts, each showing a source coloured with a heatmap. Almost all cutouts show extended structures, such as arcs, filaments, diffuse glows, jets or lobes, rather than single points. Blue frames surround 53 images and are concentrated in the upper rows. Most cutouts in the lower rows are unframed; many of these show radio galaxies with jets, lobes or bent tails. Two cutouts have green frames, the ninth cutout in the fourth row and the fifth cutout in the bottom row; both show straight parallel streaks rather than a source. A short cyan bar in the lower-left corner of each cutout marks the beam size.
\end{minipage}
\end{figure*}

We manually inspected each extracted cutout and visually compared it to the sources listed in \cite{kolokythas2025} to identify matches. We refer to the matched cutouts as \textit{tracers} throughout this work, as they allow us to assess pipeline performance. A cutout was assigned as a tracer if it spatially overlapped with an HLC source, even when the island centroid was dominated by another source. Using this procedure, we identified 122 tracers in the high-resolution dataset and 139 in the convolved dataset, exceeding the 103 sources listed in the HLC. This increase arises because a single diffuse structure can be fragmented into multiple PyBDSF islands, each producing a separate cutout. Abell 3667 \citep{2022A&A...659A.146D} provides a clear example of both this fragmentation and the overlap criterion. As seen in \autoref{fig:top 100 high res concatenated} (specifically the cutouts at row 4, column 1, and row 5, column 1), the diffuse emission in this complex system is fragmented into multiple smaller cutouts. Additionally, the second cutout is centred on a head-tail radio galaxy but also includes part of the SE relic. This cutout is therefore flagged as a tracer because it includes part of the relic, even though the island centroid is dominated by the head-tail galaxy. We deliberately retain the multiple views of the same source as these different perspectives provide distinct morphological information that the machine learning algorithm can learn from.

%------------------------------------------------%
%------------------------------------------------%

In addition, some HLC sources were not detected using the default parameters of PyBDSF, and are shown in \autoref{tab:sources_not_found}. Visual inspection of these undetected sources confirmed they have low signal-to-noise ratios that fall below the default detection thresholds.

The tracers serve a dual purpose in our pipeline: they act as known examples that guide the algorithm during training, and simultaneously provide the basis for evaluating performance. This emulates a real-world scenario where the true number of diffuse sources is unknown and differs from supervised learning approaches, where labelled data is separated into training and evaluation sets, or unsupervised methods, which use labels only for evaluation. The use of tracers for both guiding the algorithm and evaluating performance, along with the reasoning behind this approach, are discussed in \autoref{sec:guiding active learning with tracers and expertise}.

%
%% ====== Subsection
\subsection{Beam Cut Application}
\label{sec:beam cut application}
%% ===============
%

With the number of tracers in each dataset identified, we investigated the effect of different data cuts on the relevant dataset. Unlike \citet{lochner2024}, which used PyBDSF Gaussian components as complexity proxies, we apply a beam-size based cut designed to preferentially retain spatially extended emission. This removes most of the compact sources that dominate the catalogues, focusing the analysis on diffuse structures while also reducing the overall dataset size.

Assuming a Gaussian beam, we calculate the beam solid angle as \( A \approx 1.133 \times (\textrm{FWHM})^{2} \), where FWHM is the Full Width Half Maximum of the MeerKAT synthesized beam. This area is converted to pixel units and used as a reference scale. For each PyBDSF island, the measured pixel area is compared to multiples of the beam area, and sources smaller than a chosen threshold are excluded.

%------------------------------------------------%
\begin{figure*}
\centering
\begin{minipage}{\textwidth} 
    \centering
    \includegraphics[width=\textwidth]{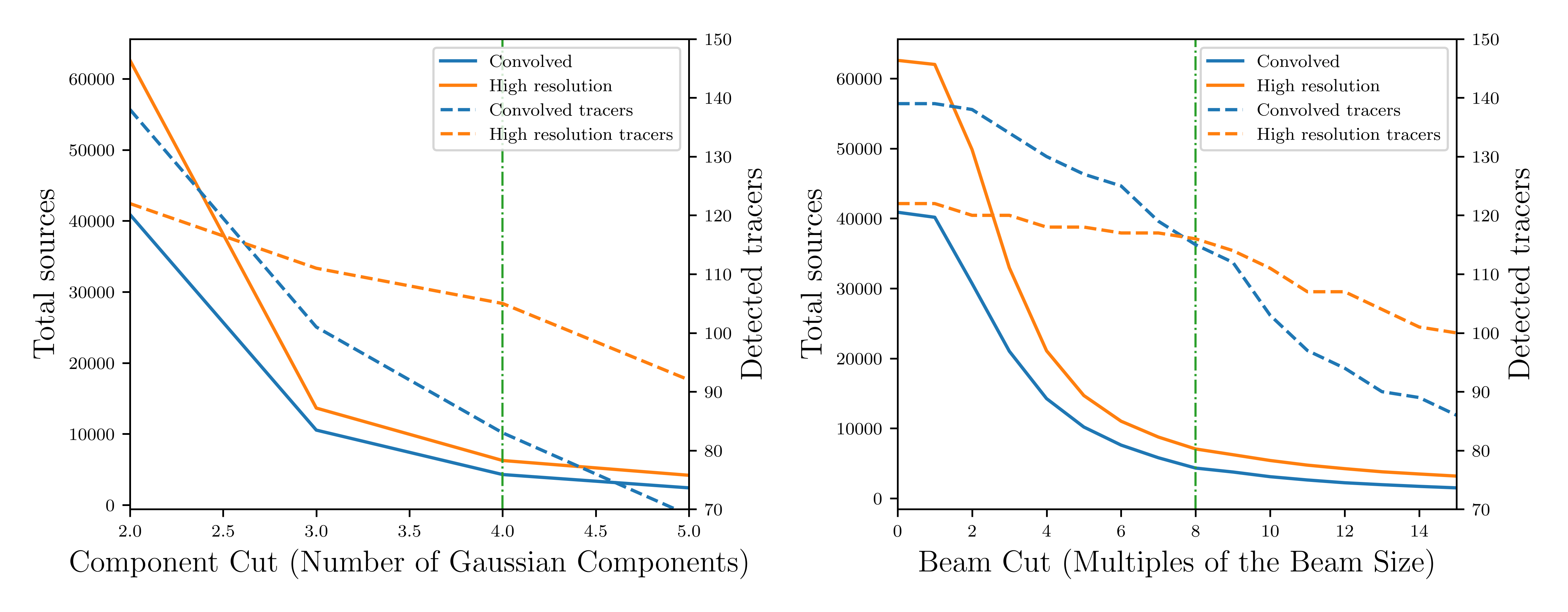}
    \caption{Comparison of data cuts applied to MGCLS sources. Left: Gaussian component-based cut from \citet{lochner2024}. Right: Beam-size cut proposed here. Plots show the number of sources (solid lines) and tracers (dashed lines) remaining in high-resolution (orange) and convolved (blue) versus cut threshold (Gaussian components or beam-size multiples). The vertical lines mark the selected thresholds. It is evident that the beam-size cut retains more tracers for similar subset sizes.}
    \label{fig:cut_comparison}
    Alt text: Two line plots side by side. Each shows the total number of sources (left axis, 0 to over 60,000) and the number of tracers kept (right axis, 70 to 150) as the cut becomes stricter. Left: the horizontal axis is the number of Gaussian components, from 2 to 5, with the chosen cut of 4 marked by a vertical dash-dotted line. Source counts drop sharply between 2 and 3 components, and tracer counts fall steadily across the whole range, most steeply for the convolved data. Right: the horizontal axis is the minimum source size in multiples of the beam size, from 0 to 15, with the chosen cut of 8 marked by a vertical dash-dotted line. Source counts drop sharply up to about 5 beams and then level off, while tracer counts fall only slowly; at the chosen cut, both tracer curves are near 115, well above the values at the chosen cut in the left panel.
\end{minipage}
\end{figure*}
%------------------------------------------------%

\autoref{fig:cut_comparison} compares the Gaussian component cut, defined as a threshold on the number of Gaussian components fitted to each source as implemented in \cite{lochner2024}, with our beam-size cut, plotting the number of sources and tracers remaining as a function of the applied cut. The key advantage of the beam-size cut becomes apparent when comparing tracer retention: for a similar final dataset size, the beam-size based cut retained a significantly higher number of tracers compared to the Gaussian component cut. This improvement occurs because the beam-size cut directly measures spatial extent rather than model complexity. Diffuse structures may have low Gaussian component counts if PyBDSF fits them with a few components only, causing them to be incorrectly excluded by component-based cuts. By measuring sizes relative to the beam, our approach better preserves extended emission regardless of its model representation.

We note that the selection of this threshold makes use of tracer information and therefore represents an informed choice rather than a fully blind selection. The implications of this are discussed in \autoref{sec:guiding active learning with tracers and expertise}.

After applying the beam-size cut, the high-resolution dataset retained 7,051 sources (116 tracers) from the initial 62,587 sources (122 tracers), and the convolved dataset retained 4,319 sources (115 tracers) from the initial 40,879 sources (139 tracers). The chosen threshold, shown as a vertical green line in \autoref{fig:cut_comparison}, was selected to balance tracer retention with dataset size reduction. The tracers excluded by this cut are primarily compact or marginally resolved sources such as compact mini-halos. While this cut does sacrifice some completeness, it substantially reduces the dataset size while retaining the majority of the tracers that are the primary focus of this study.

%
%%
%%% ====== Section
\section{Feature Extraction and Visualisation}
\label{sec:feature extraction and visualisation}
%%% ===============
%%
%

Images often contain too much information for direct pattern recognition. Even a 128×128 pixel cutout represents a high-dimensional space, making it computationally expensive and statistically challenging to identify similarities between sources. We need to reduce these images into compact numerical representations that preserve morphological information relevant to diffuse emission while discarding irrelevant details like noise and background variations. We use BYOL, a self-supervised deep learning algorithm, to extract these representations automatically without requiring labelled training data.

%
%% ====== Subsection
\subsection{Preprocessing}
\label{sec:preprocessing}
%% ===============
%

Cutouts whose bounding boxes intersect the edge of the original MGCLS field are removed, as truncated images could bias feature extraction. For each remaining PyBDSF island, we define a bounding box enlarged by a factor of 1.15 to ensure the full source is captured, and resize the resulting cutout to 128×128 pixels using \textsc{skimage} \citep{van_der_walt_2014}, with smaller sources interpolated using bicubic interpolation and larger sources downsampled.

Following \citet{Lochner2021}, we apply sigma clipping and masking using the process implemented in \textsc{Astronomaly}. Pixels outside the source contour, determined with \textsc{OpenCv}, are masked if they lie outside a $3\sigma$ threshold, where $\sigma$ is the standard deviation of the local background noise estimated from the pixel intensity distribution. This isolates genuine astronomical emission while removing background noise. For visualisation only, \textit{asinh} scaling is applied to adjust brightness ranges without affecting the data used for feature extraction.

%
%% ====== Subsection
\subsection{BYOL}
\label{sec:feature extraction byol}
%% ===============
%

Bootstrap Your Own Latent (BYOL, \citealp{grill2020bootstrap}) is a self-supervised learning algorithm that learns low-dimensional representations from unlabelled images. Unlike contrastive methods that require negative sample pairs, BYOL uses two neural networks, an online network and a target network, that learn from each other through an iterative prediction task. Two augmented views of the same image are generated; the online network predicts the representation of the target network of one view, while the weights of the target network are updated via a slow-moving average of the weights of the online network. This bootstrapping process allows progressive improvement without requiring labelled data or explicit negative examples.

We implement BYOL using the EfficientNet-B0 architecture \citep{EfficientNet}, chosen for its balance between performance and computational efficiency. The network outputs a 1280-dimensional feature vector for each image. Following \citet{lochner2024}, we train for 100 epochs using the Adam optimiser, with the same hyperparameter values and training setup. Training and validation loss curves (\autoref{sec:model validation and loss curves}) confirm that the model learns robust representations without overfitting.

%
%% ====== Subsection
\subsection{Dimensionality Reduction}
\label{sec:dimensionality reduction}
%% ===============
%

To maximise performance and computational efficiency, we applied Principal Component Analysis (PCA, \citealp{hotelling1933pca}) to reduce the dimensionality of the BYOL feature space while retaining 95\% of the variance. This reduced the representation to 37 components for the convolved data and 40 for the high-resolution data, lowering data volumes by over 97\%. PCA removes redundancy by eliminating correlated features and often improves downstream performance by reducing noise. Although it assumes linear relationships between features and may not capture all structure in the BYOL embedding space, PCA remains a widely used and effective approach in similar applications.

%
%% ====== Subsection
\subsection{Visualisation}
\label{sec:visualisation}
%% ===============
%

Visualising the feature space provides insight into how well the model learned meaningful representations and how different source types are organised relative to one another. However, the PCA-reduced feature space remains too high-dimensional for direct visualisation. To overcome this, we use Uniform Manifold Approximation and Projection (UMAP, \citealp{UMAP2020}) to embed the data into two dimensions while preserving local and global relationships between points. We implemented UMAP via the \texttt{umap-learn} Python package \citep{UMAP2021_python} with a minimum distance of 0.1 and 20 neighbours, which produced stable embeddings.

The resulting embedding places similar sources close together, enabling intuitive exploration of structure in the feature space. Due to the random nature of UMAP and the sensitivity to parameter choices, these embeddings are used for visualisation only; all quantitative analyses and evaluations are performed on the PCA-reduced features. While UMAP should not be interpreted as an exact projection, it reliably reflects genuine similarities learned by BYOL when the input features are well structured.

%
%%
%%% ====== Section
\section{Active Learning with Protege}
\label{sec:active learning with protege}
%%% ===============
%%
%

\citet{Walmsley2022b} and \citet{Lochner2021} demonstrated that deep features, while effective at grouping together similar sources, produce relatively uniform spaces in which rare or scientifically interesting sources are often missed by traditional anomaly detection methods. To address this, they proposed replacing anomaly scoring with a regression-based approach using Gaussian processes \citep{rasmussen2006gaussian} combined with active learning. This allows users to iteratively refine searches based on their specific scientific interests rather than generic outlier metrics. This approach has been incorporated into the \textsc{Astronomaly} framework as \textsc{Protege} \citep{lochner2024}. We use BYOL-extracted features (\autoref{sec:feature extraction byol}) as input to \textsc{Protege} to identify diffuse emission candidates.

\textsc{Protege} operates through iterative active learning. In each iteration, the algorithm uses an acquisition function to select a batch of sources for human inspection, prioritising those that are expected to most improve its understanding of user preferences. The user labels these sources, and the algorithm updates its model to better predict which unlabelled sources match the interests of the user (see \autoref{sec:illustration of algorithm improvement} for an illustration). This process continues until the improvement in ranking performance from additional labels becomes negligible, or until the user is satisfied with the ranked recommendations.

We adapt the implementation of \textsc{Protege} in three ways. First, we focus exclusively on diffuse cluster emission rather than multiple anomaly types. Second, we simplify the labelling scheme: sources matching the tracers (\autoref{sec:tracers}) receive a score of 5, while all others receive 0 (see \autoref{fig:examples_of_tracers}). This binary approach reduces labelling inconsistency compared to the full 0-5 gradient scale. Third, we increase the batch size labelled in each iteration from 10 to 15 labels per iteration (compared to \citealt{lochner2024}), as preliminary tests showed 10 labels were insufficient to quickly identify the first instances of tracers. We limit the total number of labelled sources to 300 across all iterations (20 iterations total).

The simplified labelling scheme trades flexibility for consistency. While the full gradient allows expressing degrees of interest, it introduces subjectivity in how intermediate scores are assigned. Our binary approach provides a controlled test of whether \textsc{Protege} can learn to distinguish a specific morphological class from the general population when given clear examples.

Detailed computational requirements, including hardware specifications and runtimes for each pipeline stage, are provided in \autoref{sec:compute}.

%
%%
%%% ====== Section
\section{Results}
\label{sec:results}
%%% ===============
%%
%

Evaluating any non-supervised learning approach presents challenges, particularly for any dataset where the full population of targets is unknown, as is the case when exploring new data. In this work, tracers are used both to guide the iterative active learning process and to evaluate performance, where a fraction of them will have been seen by the algorithm during training. This reflects the design of \textsc{Protege} rather than a standard supervised approach.

Our approach generates ``cumulative curves'' (or ``cumulative anomaly curves'') that track how the tracers rank as the algorithm learns during active learning. At each iteration, \textsc{Protege} assigns scores to all sources based on predicted ``interestingness,'' producing a ranked list with the most promising candidates at the top. These curves show the cumulative number of tracers recovered within the top \textit{N} ranked sources, representing the reality that a user only has the capacity to inspect a limited number of sources. Steeper curves indicate faster recovery of tracers, while a diagonal curve reflects random ordering.

In practice, without a ground truth dataset, performance will be evaluated largely by a manual and visual inspection of the top \textit{N} ranked candidates, such as \autoref{fig:top 100 high res concatenated} where the top 100 are shown.

%
%% ====== Subsection
\subsection{High Resolution versus Convolved}
\label{sec:high resolution versus convolved}
%% ===============
%

We analysed both MGCLS datasets separately to assess how image resolution affects detection and feature representation. Diffuse cluster emission spans multiple spatial scales, and different resolutions reveal complementary information: high-resolution imaging captures fine structural detail but can fragment or suppress faint extended emission, while convolved images enhance sensitivity to low-surface-brightness structures at the cost of fine detail. Analysing both allows us to evaluate how these differences influence feature extraction, source ranking, and overall detection performance.

Both datasets underwent preprocessing (\autoref{sec:preprocessing}), feature extraction (\autoref{sec:feature extraction byol}), dimensionality reduction (\autoref{sec:dimensionality reduction}), and analysis with \textsc{Protege} (\autoref{sec:active learning with protege}), with binary labelling applied to tracers and non-tracers.

%------------------------------------------------%
\begin{figure*}
    \begin{minipage}{\textwidth}
        \centering
            \includegraphics[width=0.4975\textwidth]{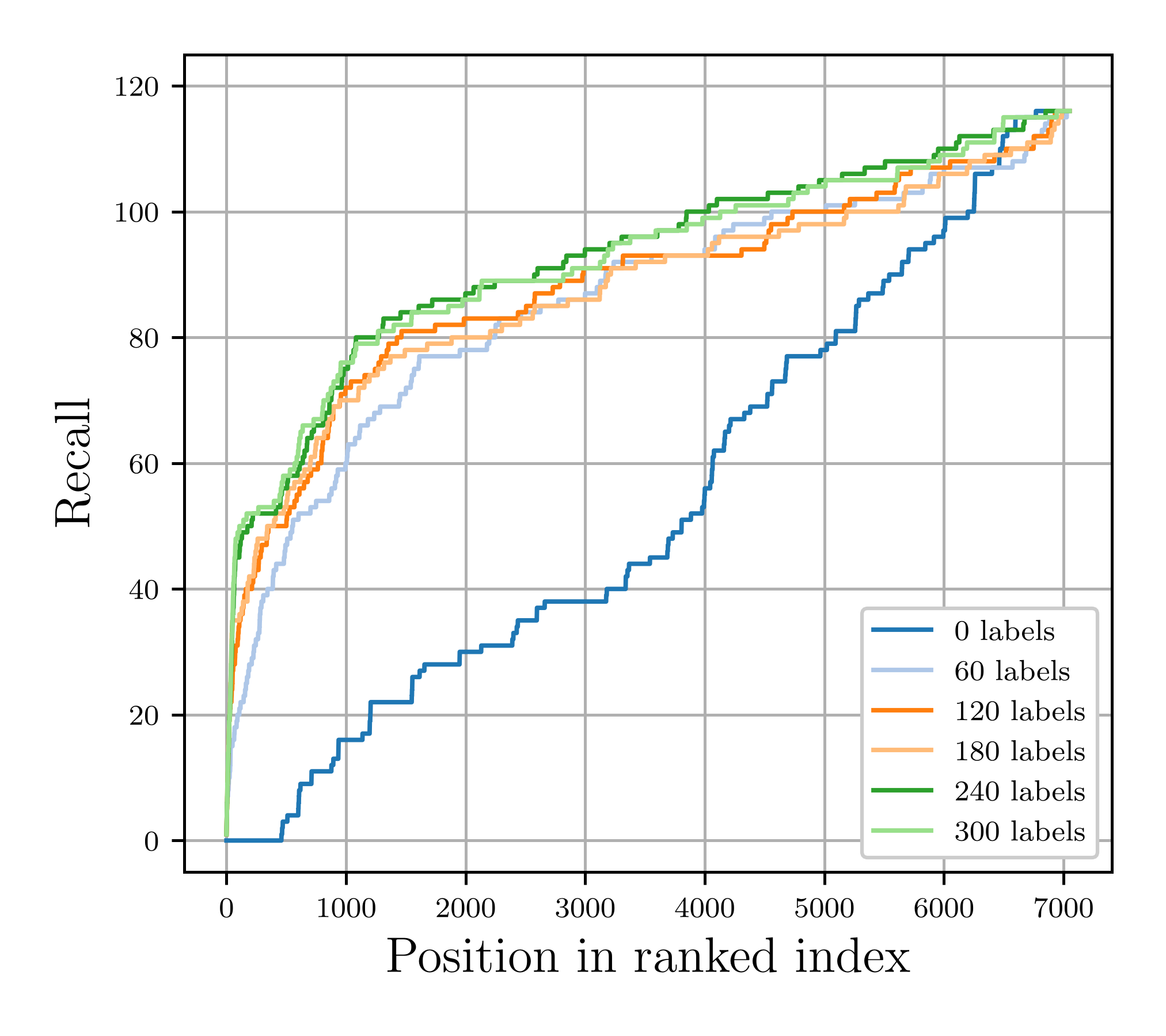} \hfill
            \includegraphics[width=0.4975\textwidth]{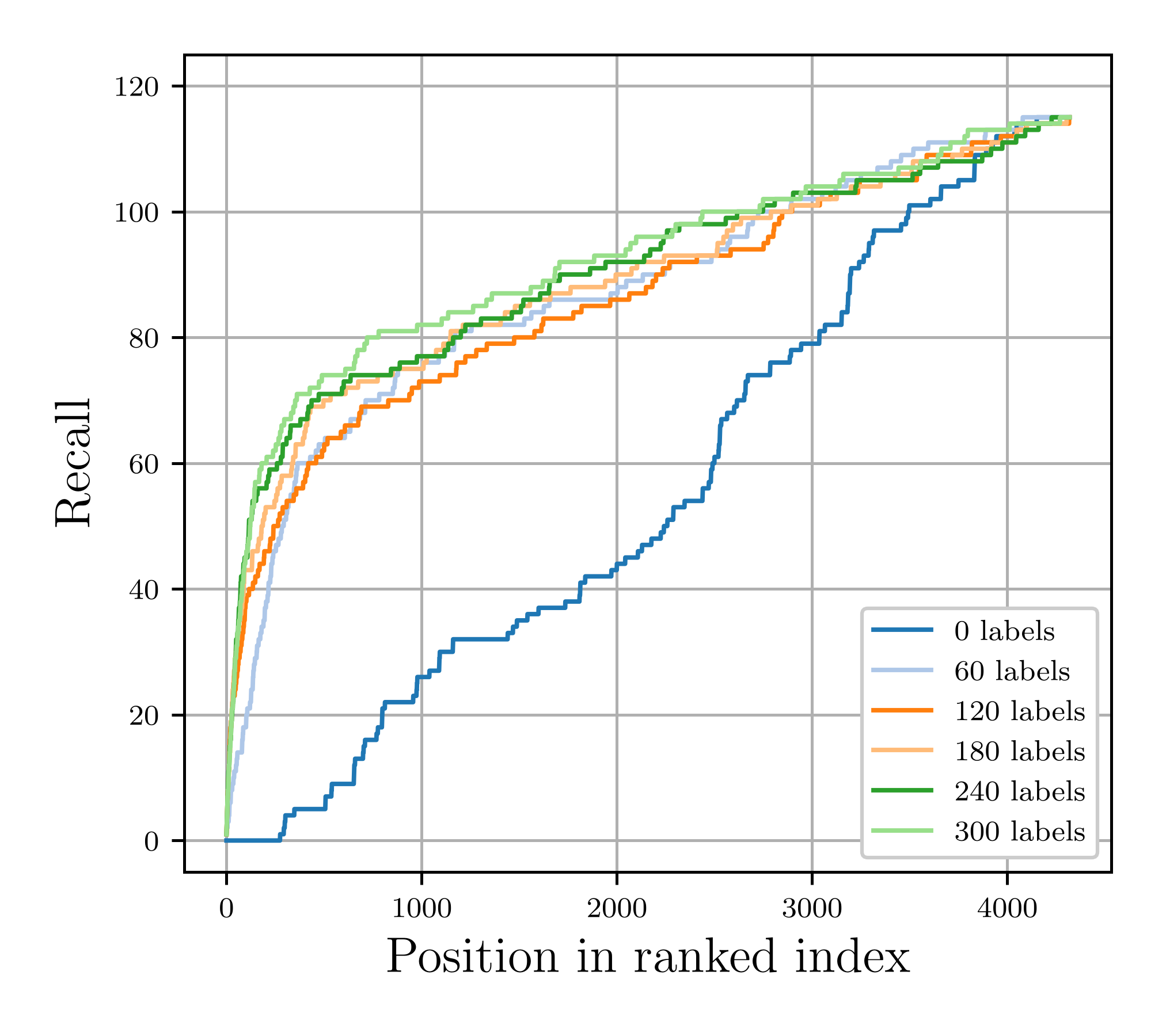}
        \caption{Cumulative anomaly curves for the high-resolution dataset (on the left with 116 tracers in total) and the convolved dataset (on the right with 115 tracers in total), showing the cumulative number of tracers recovered within the ranked list for different labelling iterations. The ``$0$ labels'' curve corresponds to the initial distribution of tracers prior to labelling. Both datasets show significant improvements in the cumulative number of tracers with iterative labelling, although diminishing returns are observed in later iterations.}
        \label{fig:recall_subset_high_vs_convolved_b}
        Alt text: Two step-like line plots of recall (number of tracers recovered, 0 to 120) against position in the ranked list (0 to about 7,000 on the left, 0 to about 4,300 on the right). Each panel has six curves, for 0, 60, 120, 180, 240 and 300 labels. The 0-labels curve rises slowly and fairly evenly along the whole list. All the labelled curves rise sharply at the very start of the list, then climb gradually to the full total at the end. The labelled curves lie close together, showing little gain after the first 60 labels.
    \end{minipage}
\end{figure*}
%------------------------------------------------%

\autoref{fig:recall_subset_high_vs_convolved_b} shows the cumulative anomaly curves across several labelling iterations for the different resolutions. \autoref{fig:recall_subset_high_vs_convolved_b} shows that roughly 42\% (49 of 116 for the high resolution) of the tracers can be recovered within 1\% (top 100) of the subset with minimal human labelling (300 labels), highlighting the strong performance of \textsc{Protege} in prioritising diffuse emission candidates. Significant improvements are also seen with as few as 60 labels provided (with 50\% of the tracers recovered in 15\% of the data), and further iterations continue to improve the rankings, although with diminishing returns. This trend matches previous findings \citep{Etsebeth2024}, where improvements plateau once sufficient examples are provided.

Comparing the two datasets, the cumulative curve for the convolved images rises to a higher value at the top of the ranked list (top 1000), suggesting better early recovery of tracers. This likely reflects that convolved images enhance extended low-surface-brightness structures while smoothing fine-scale noise, making diffuse emission more distinguishable in the feature space. However, differences in the total number of sources between datasets, and the choice of top \textit{N} sources inspected, must be taken into account before any comparison can be made. As explored in \autoref{sec:overall comparison}, examining the top 500 candidates allows a fairer comparison of pipeline performance across resolutions, mimicking a realistic scenario where only a subset of sources in a new dataset can be inspected.

Overall, both datasets demonstrate that \textsc{Protege} efficiently prioritises diffuse emission candidates, with early detection and reduced human inspection workload achieved across both resolutions. 

%------------------------------------------------%
\begin{figure*}
    \begin{minipage}{\textwidth}
        \centering
            \includegraphics[width=0.475\textwidth]{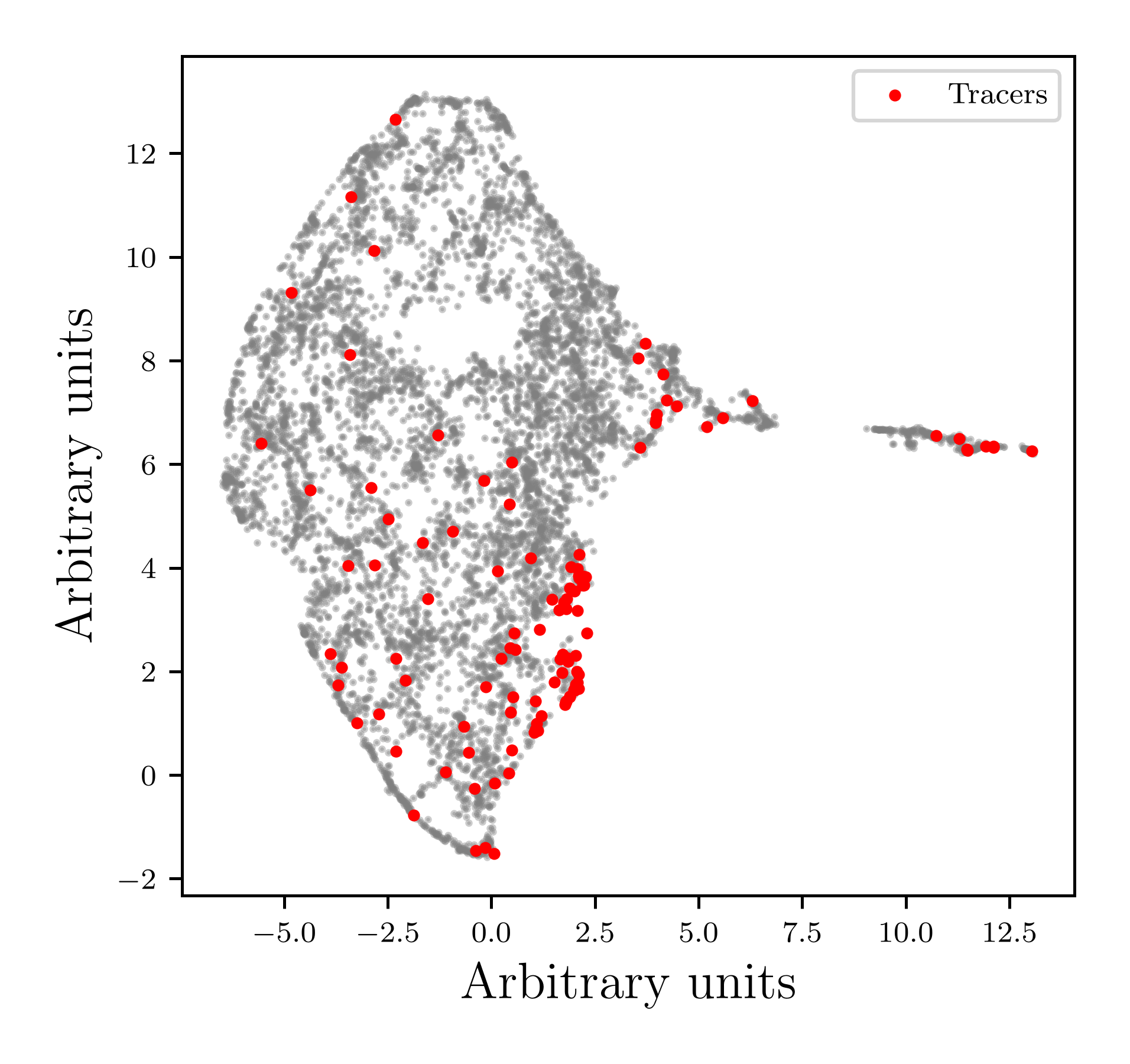} \hfill
            \includegraphics[width=0.475\textwidth]{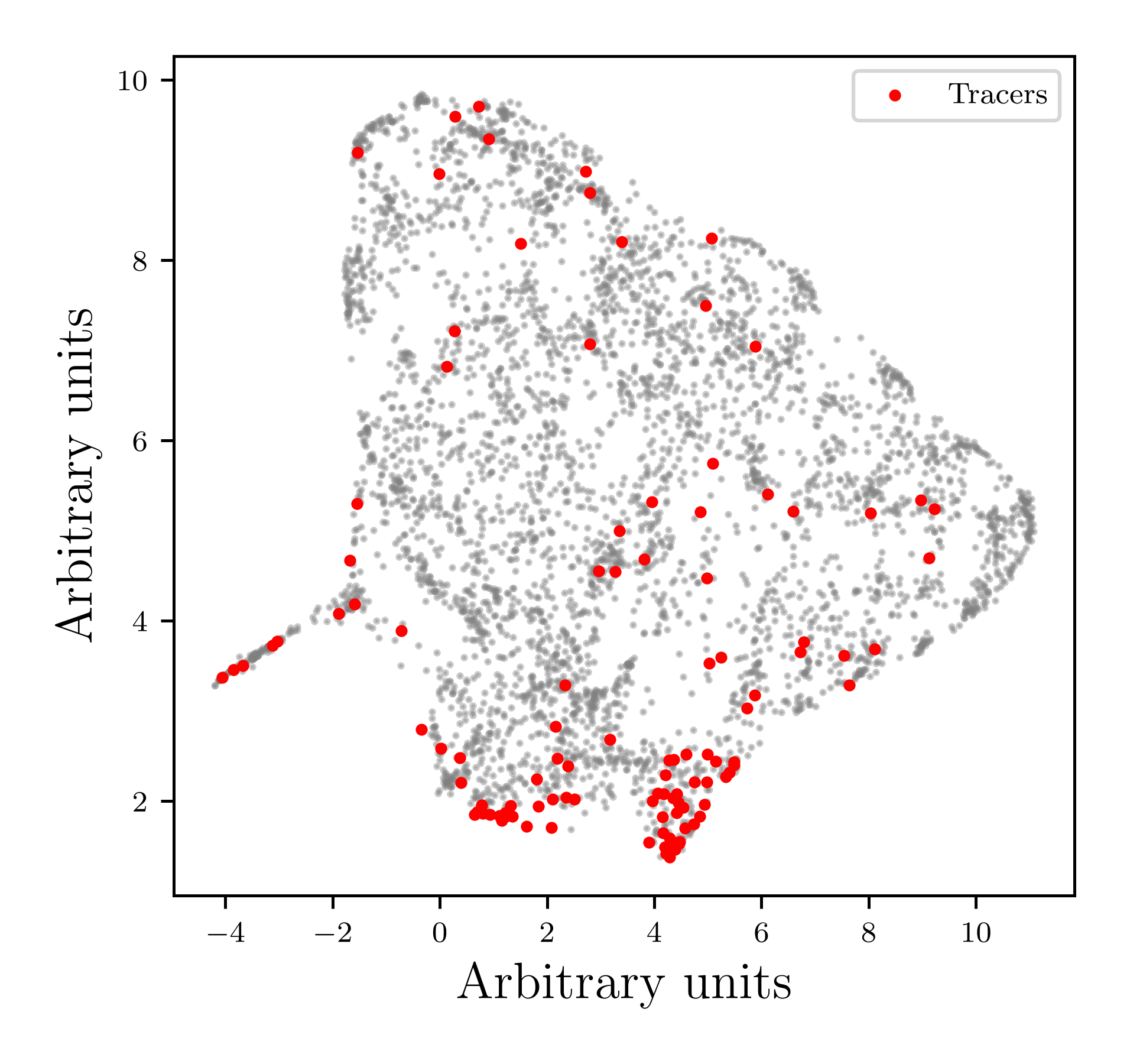}
        \caption{UMAP visualisations of the high resolution (left) and convolved (right) feature spaces. Red points indicate the tracers and grey points represent all other sources. In both cases, the tracers form several groupings with some sources dispersed across the feature space, presenting a challenge for detection with machine learning.}
        \label{fig:UMAP of the high and convolved subsets}
        Alt text: Two scatter plots in arbitrary units, each with thousands of grey points (all sources except the tracers) and red points (tracers) on top. Left, high resolution: the points form one large group (cluster) with a narrow arm extending to the right and a small separate group at the far right. Red points gather in a dense group near the lower centre of the cluster and a smaller group where the arm begins, with a few in the separate group and the rest scattered across the cluster. Right, convolved: the points form one large cluster with a narrow arm extending to the lower left. Red points gather in two dense groups along the bottom edge and a few along the arm, with the rest scattered across the cluster.
    \end{minipage}
\end{figure*}
%------------------------------------------------%

\autoref{fig:UMAP of the high and convolved subsets} visualises the feature spaces of both datasets using UMAP, with tracers in red and all other sources in grey. The disjointed groupings and dispersion of tracers across the feature space reflects the diversity of the learned representations in the two datasets. This scattered distribution poses two challenges: the algorithm must learn multiple distinct morphological patterns rather than a single coherent cluster, and the isolated tracers embedded among the general population are likely to be missed, as active learning prioritises regions similar to already-labelled examples. The structure of the feature space is further discussed in \autoref{sec:feature space structure}.

%
%% ====== Subsection
\subsection{Concatenated Features}
\label{sec:concatenated features}
%% ===============
%

%------------------------------------------------%
\begin{figure}
\begin{minipage}{\columnwidth}
    \centering
        \includegraphics[width=\columnwidth]{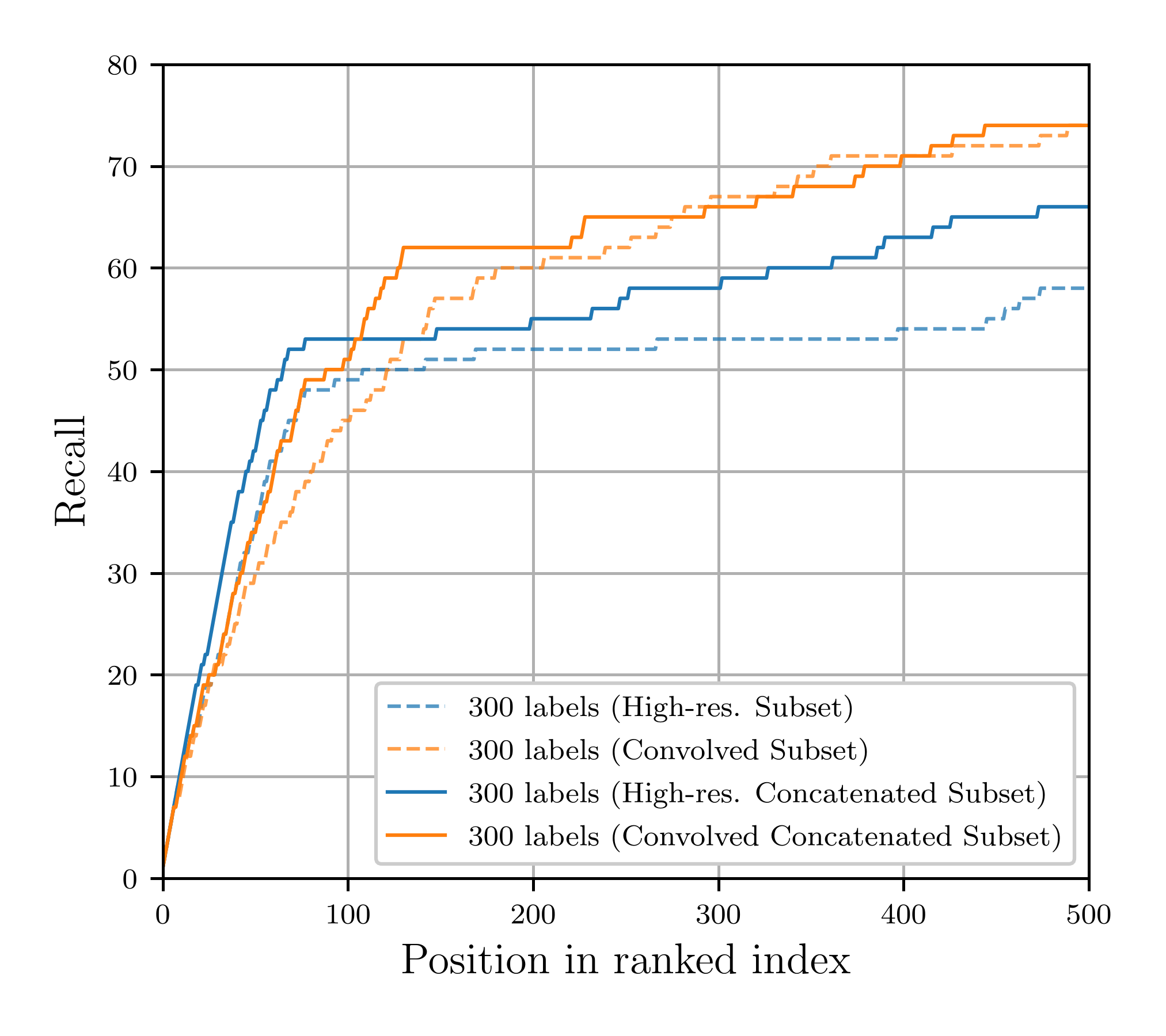}
    \caption{Cumulative curves for the top 500 sources across all configurations: standalone high-resolution and convolved datasets, plus concatenated features using either high-resolution or convolved coordinates.}
    \label{fig:recall_previous_four}
    Alt text: Step-like line plot of recall (number of tracers recovered, 0 to 80) against position in the ranked list (0 to 500) for four configurations after 300 labels. Blue lines use high-resolution coordinates and orange lines use convolved coordinates; dashed lines are single-resolution features and solid lines are concatenated features. The two blue curves rise fastest, reaching about 50 tracers by position 100, then flatten, ending near 58 (dashed) and 66 (solid). The two orange curves rise more slowly at first but keep climbing, both ending near 74. At position 100, each solid curve lies above its dashed counterpart.
\end{minipage}
\end{figure}
%------------------------------------------------%

As shown in \autoref{fig:recall_previous_four}, the high-resolution and the convolved datasets exhibit different cumulative recall behaviours within the top 500 ranked sources. High-resolution features yield a steeper initial rise, suggesting that known tracers are concentrated near the top of the list. Conversely, convolved features exhibit a more gradual increase but recover a larger fraction of tracers across the full top 500. Because these behaviours are complementary, we investigated combining both resolutions into a single representation by simply concatenating the two sets of features before dimensionality reduction.

To ensure consistency, we used PyBDSF island coordinates and size information to generate matching cutouts across resolutions. By applying coordinates from one dataset to images from the other, we extracted BYOL features consistently from both resolutions. Each source was thus represented by two 1280-dimensional vectors, which were concatenated into a single 2560-dimensional vector before applying PCA.

We tested two configurations: (1) using high-resolution coordinates on convolved images, and (2) using convolved coordinates on high-resolution images. Because PyBDSF identifies different numbers and extents of islands in each dataset, these configurations produce distinct source samples even for the same fields. 

The cumulative recall performance of these concatenated representations (solid lines in \autoref{fig:recall_previous_four}) was evaluated using 300 labelled sources over 20 active learning iterations, focusing on the top 500 ranked sources to emphasise early detection.

%
%% ====== Subsection
\subsection{Overall Comparison}
\label{sec:overall comparison}
%% ===============
%

\autoref{fig:recall_previous_four} compares the performance of all four configurations: the standalone high-resolution and convolved features, along with the two concatenated representations. Across all configurations, high-resolution features consistently exhibit steeper initial increases in cumulative recall within the top 100 sources. 

Specifically, 49 out of 116 tracers appear in the top 100 for the high-resolution features, compared to 53 for the high-resolution concatenated features. This indicates that high-resolution imaging alone is sufficient for targeted searches and rapid identification of the most morphologically distinct diffuse emission, with the addition of convolved information producing only a minor improvement in the top 100. Beyond the top 100 however, including convolved information improves overall recovery: in the top 500 sources, the high-resolution dataset alone recovers 58 out of 116 tracers, while the high-resolution concatenated representation recovers 66 out of 116.

For the convolved features, the pattern is reversed. In the top 500, both the convolved dataset and the convolved concatenated dataset recovers 74 out of 115 tracers, showing no improvement with the added high-resolution information. However, in the top 100, the convolved features recover 45 tracers, whereas the convolved concatenated features recover 51, demonstrating that the high-resolution information enhances early detection rates.

These trends reflect the complementary strengths of the two resolutions. High-resolution features excel at identifying the most prominent and morphologically distinct sources, while convolved features improve recovery of fainter, extended emission across larger scales. Concatenated representations combine these advantages, retaining early detection performance from the high-resolution features and improved overall recovery from the convolved features. Consequently, when both resolutions are available, concatenated representations enhance both early detection and overall recovery, whereas the standalone high-resolution dataset remains suitable for rapid identification of prominent sources and the standalone convolved dataset is optimal for comprehensive searches prioritising completeness.

%
%% ====== Subsection
\subsection{Top 100 Investigation}
\label{sec:top 100 investigation}
%% ===============
%

We conducted a visual inspection of the top 100 ranked sources from the concatenated high-resolution coordinate subset, which recovered the most tracers within this range. \autoref{fig:top 100 high res concatenated} displays these sources, with known tracers highlighted by blue frames.

The visual inspection reveals results that the cumulative plots do not reflect: 98 of the 100 top-ranked sources exhibit diffuse emission of some sort. Of these, 53 correspond to diffuse cluster emission and match the known tracers. The remaining 45 sources include radio AGN galaxies with jets and/or extended lobes, radio AGN galaxies with unusual structures (such as bent tail or wide angle tail morphologies), face-on spiral galaxies where the radio emission traces star-forming regions, and several unidentified extended structures. All display some form of diffuse radio emission, albeit produced by different physical mechanisms. The other 2 sources are imaging artefacts exhibiting radial spikes, as highlighted by the green frames in \autoref{fig:top 100 high res concatenated}.

These sources exhibit diffuse radio structures and could represent potential cluster-associated emission. However, confirming this would require additional investigation, such as X-ray imaging to detect associated intra-cluster medium or spectroscopic redshifts to establish cluster membership. Some sources may instead correspond to background galaxies, diffuse emission from AGN in galaxy groups rather than cluster-associated emission, or radio galaxies with extended lobes viewed from unusual angles.

While the algorithm successfully identifies diffuse sources, distinguishing genuine cluster-associated diffuse emission from other instances requires a follow-up analysis beyond the scope of this automated detection pipeline.

%
%% ====== Subsection
\subsection{Investigation of Tracers in Lower-Ranked Positions}
\label{sec:investigation of tracers in lower-ranked positions}
%% ===============
%

%------------------------------------------------%
\begin{figure*}
\begin{minipage}{\textwidth} 
    \centering
    \includegraphics[width=0.975\textwidth]{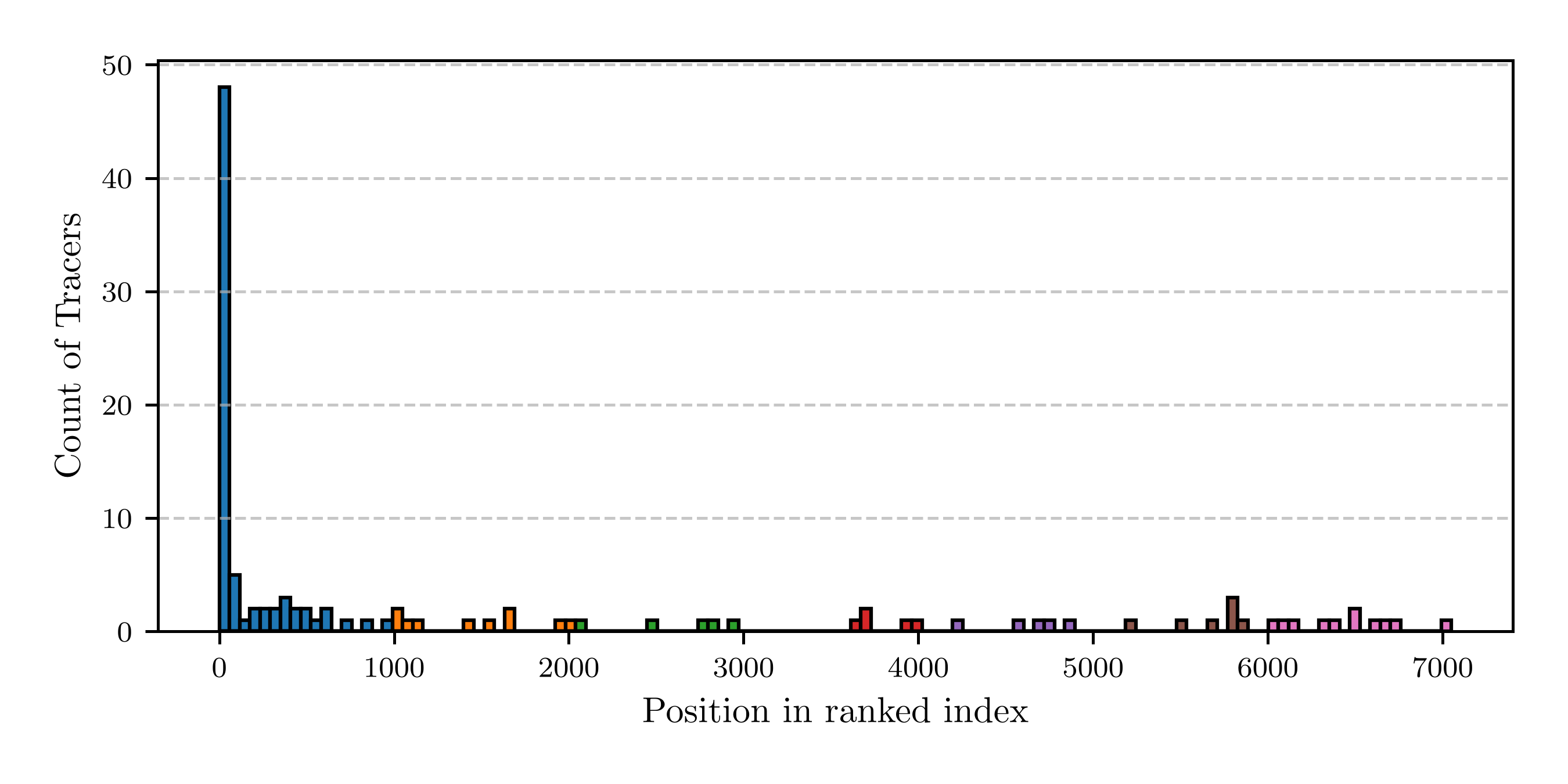}
    \caption{Histogram of tracers across the full ranked list using the concatenated high-resolution features. Bars are colour-coded in bins of 1000 sources each. Most tracers appear within the top 1000 sources, but some rank considerably lower.}
    \label{fig:histogram of tracers}
    Alt text: Histogram of the number of tracers against position in the ranked list (0 to about 7,000). A single very tall bar at the start of the list holds nearly 50 tracers. The remaining bars are short, each holding one to five tracers, and are spread thinly along the whole list. Bar colours change every 1,000 positions, in the order blue, orange, green, red, purple, brown and pink, matching the frame colours in Figure 8.
\end{minipage}
\end{figure*}
%------------------------------------------------%

While the ranking produced by the algorithm places most tracers near the top of the list, understanding why some rank poorly provides insight into its limitations. \autoref{fig:histogram of tracers} shows their distribution across the full ranked list using the concatenated high-resolution coordinate configuration. Most are concentrated within the top 1000 sources, but the distribution extends all the way down the list.

\begin{figure*}
\begin{minipage}{\textwidth} 
    \centering
    \includegraphics[width=0.995\textwidth]{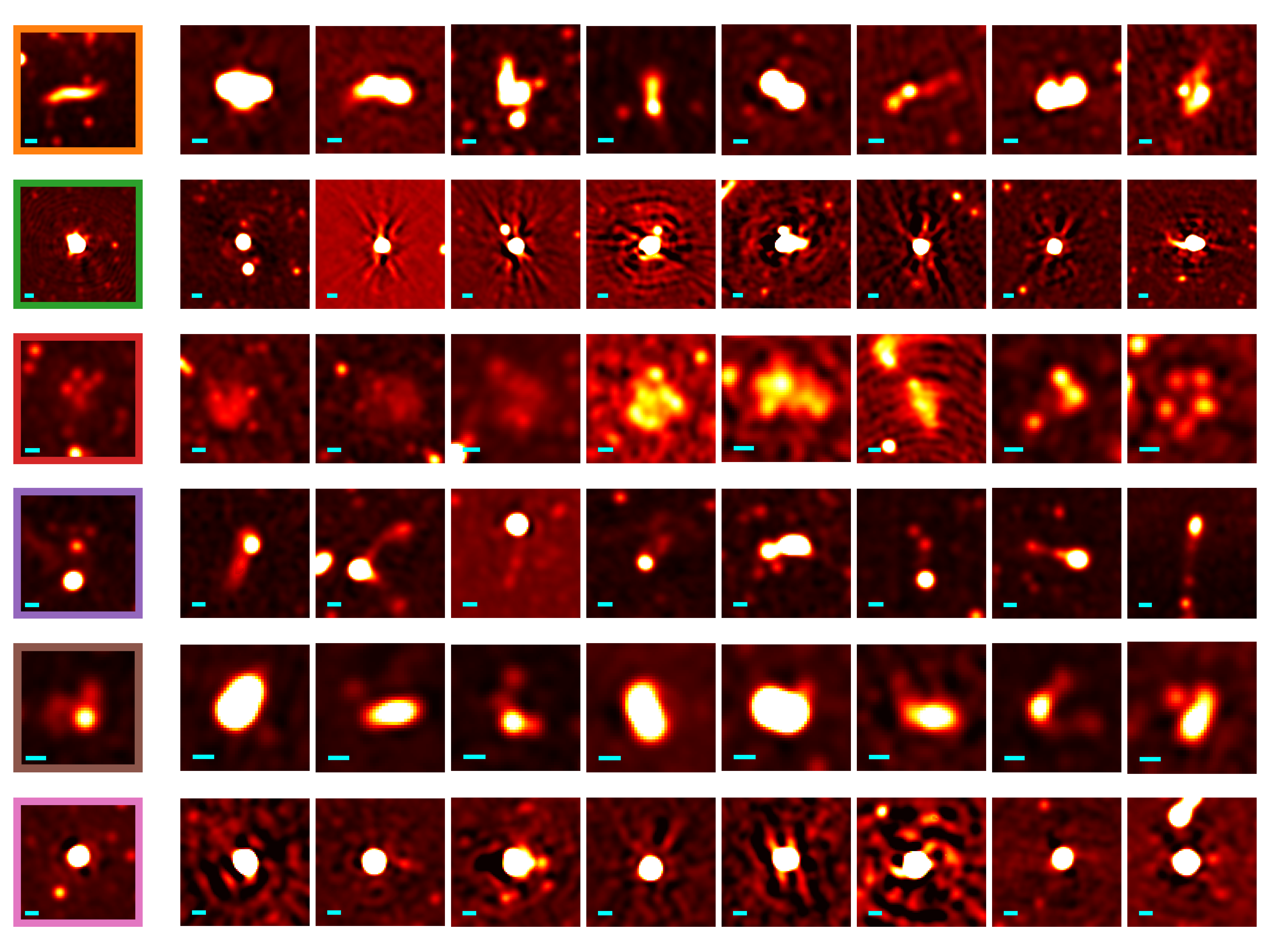}
    \caption{Tracers selected from throughout the ranked list (coloured frames matching the bins in \autoref{fig:histogram of tracers}), shown with their eight nearest neighbours. The clusters hosting the representative tracers, from top to bottom, are: J2023.4-5535, J0510.2-4519, MACS J0257, J0637.3-4828, J0145.2-6033, and J1423.7-5412. These low-ranked sources are typically faint, compact, or barely resolved, with morphologies that resemble the general population of compact sources, making them difficult to identify during active learning.}
    \label{fig:tracers throughout}
    Alt text: Six rows of nine cutouts. In each row, the first image is a low-ranked tracer, framed in the colour of its position bin (orange, green, red, purple, brown and pink from top to bottom), followed by its eight nearest neighbours. Row 1: a small, faint, elongated source, next to mostly bright compact or slightly elongated blobs. Row 2: a compact source with faint surroundings, next to bright point sources with radial spokes or rings. Row 3: a few faint clumps, next to faint patchy emission. Row 4: two compact sources with diffuse emission surrounding them, next to compact sources, some with short faint tails. Row 5: a faint compact source surrounded by faint diffuse emission, next to bright oval blobs. Row 6: another faint compact source surrounded by faint diffuse emission, next to compact sources surrounded by rippled patterns. In every row the neighbours are compact or ambiguous rather than clearly extended diffuse emission. A short cyan bar in the lower-left corner of each image marks the beam size.
\end{minipage}
\end{figure*}

To investigate why some tracers rank poorly, we partitioned the ranked list into coloured bins (each containing 1000 sources) and examined representative sources from each. \autoref{fig:tracers throughout} shows these representative tracers alongside their eight nearest neighbours in feature space (measured using Euclidean distance).

%------------------------------------------------%
%------------------------------------------------%

Visual inspection reveals why these sources challenge the algorithm. Most are small with low signal-to-noise ratios, leaving them with few distinctive morphological features that can be learned by BYOL. Their nearest neighbours similarly show compact or ambiguous morphologies, confirming that these tracers occupy regions dominated by non-diffuse sources.

\begin{figure*}
    \begin{minipage}{\textwidth}
        \centering
            
            \begin{minipage}[c]{0.31\textwidth}
                \centering
                \includegraphics[width=\textwidth]{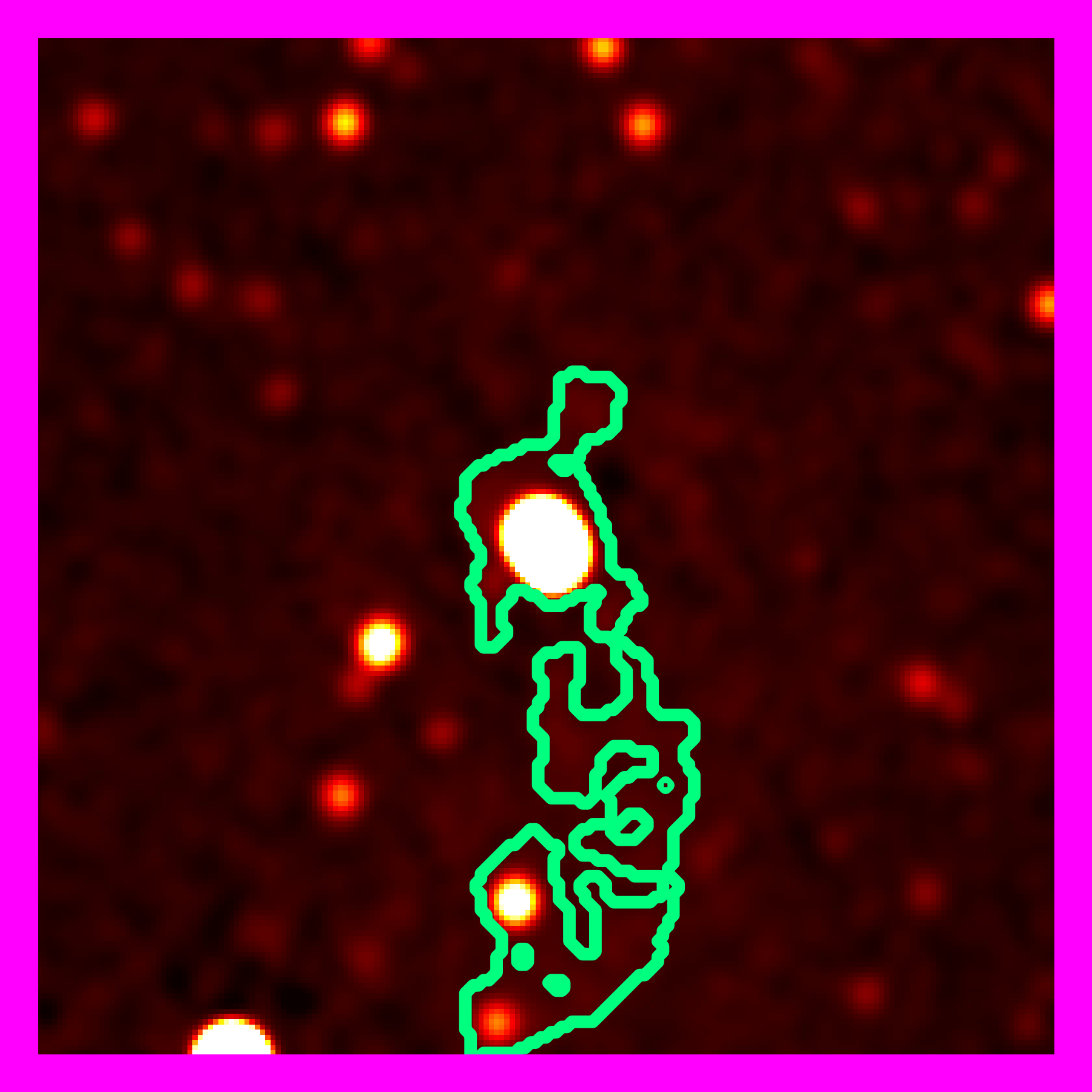}
            \end{minipage}\hfill
            \begin{minipage}[c]{0.69\textwidth}
                \centering
                \includegraphics[width=\textwidth]{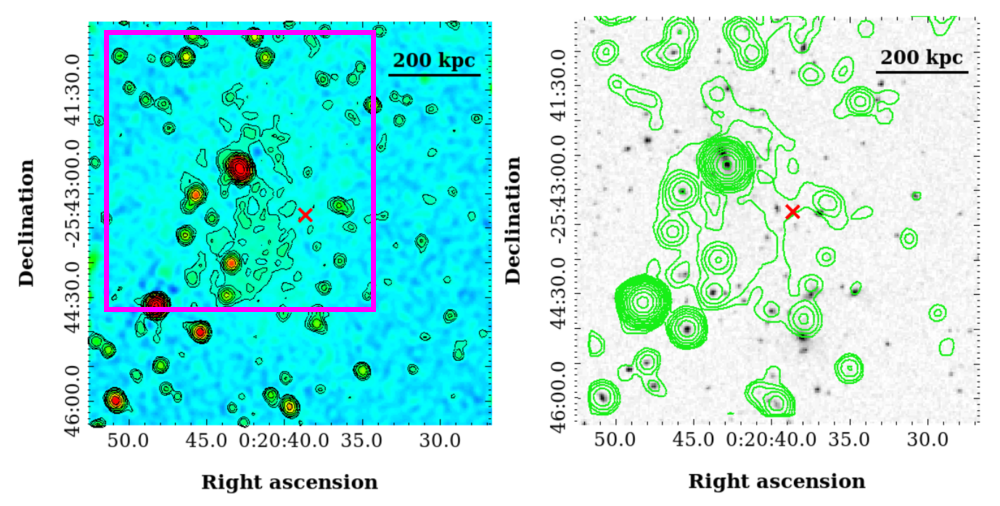}
            \end{minipage}
            \caption{Diffuse radio emission associated with the galaxy cluster Abell 22. On the left: the candidate radio halo cutout created using our processing pipeline, with the PyBDSF island boundary overlaid in green. On the middle and right: the MGCLS radio and optical view of Abell 22 adapted from \citet{kolokythas2025}, showing full-resolution (middle) and low-resolution (right) 1.28 GHz radio contours overlaid on optical imaging. The pink overlaid box on the middle panel indicates the region shown in our cutout on the left.}
            \label{fig:abell22_comparison}
            Alt text: Three panels. Left: our radio cutout coloured with a heatmap and with a pink frame, showing several bright compact sources on a dark background. A green outline marks the PyBDSF island, an irregular, vertically elongated region with a bright compact source in its upper part; the diffuse emission inside it is barely visible. Middle: a wider radio image in a blue colour scale with black contours, showing many compact sources and faint, patchy contours near the centre around a red cross. A pink box marks the area of the left cutout, and a bar shows 200 kpc. Right: the same field shown as a greyscale optical image with green low-resolution radio contours. Besides the compact sources, these contours trace a larger, continuous region of diffuse emission around the red cross. The middle and right panels have axes of right ascension and declination.
    \end{minipage}
\end{figure*}

This ambiguity is further illustrated by a candidate diffuse cluster emission source in Abell 22, shown in \autoref{fig:abell22_comparison}. In our cutout (left panel), the emission appears faint and ambiguous, whereas the MGCLS analysis (right panel) identifies it as a candidate diffuse radio halo (adapted from \citealp{kolokythas2025}). This demonstrates how easily such sources can be overlooked without additional contextual information, such as cluster centre positions or X-ray data. This highlights an inherent limitation that is further discussed in \autoref{sec:feature space structure}: morphological diversity within the ``diffuse cluster emission'' category results in some examples resembling the general population more than other tracers. 

%------------------------------------------------%
%------------------------------------------------%

This limitation is not evenly distributed across the morphological sub-classes. A further analysis of the ranked output reveals that relics are recovered much more efficiently than halos. A detailed breakdown of the class-specific rates, along with an investigation into morphology-specific training, is provided in \autoref{sec:morphology}.

Nonetheless, our pipeline with the concatenated features achieves strong overall performance. After labelling 300 sources (20 iterations of 15 labels each), 53 of the 116 tracers are detected after viewing a total of 100 sources (and half of all tracers, 58 of 116, are detected after viewing a total of 250 sources) out of the original 62,587 sources. This dramatic reduction in human effort makes the search for diffuse emission feasible in the SKA-scale era. 

%
%%
%%% ====== Section
\section{Discussion}
\label{sec:discussion}
%%% ===============
%%
%

Our results demonstrate that \textsc{Protege} can effectively identify diffuse emission in large surveys, but the methodology raises important questions about the role of labelled data, the limits of morphology-based classification, and what constitutes ``successful detection'' when targets are morphologically diverse. We address these questions in the following subsections.

%
%% ====== Subsection
\subsection{Guiding Active Learning with Tracers and Expertise}
\label{sec:guiding active learning with tracers and expertise}
%% ===============
%

Known diffuse cluster emission tracers are used both to guide the pipeline and to evaluate its effectiveness. For example, in defining the beam-size cut (\autoref{sec:beam cut application}), the tracers reveal how different thresholds affect the retention of extended structures. While both Gaussian component and beam-size cuts could be applied, examining the tracers allows us to choose a cut and a threshold that better preserves known diffuse emission sources.

Comparing Gaussian component and beam-size based cuts (\autoref{fig:cut_comparison}) highlights why this matters in practice. Some diffuse structures consist of only a few Gaussian components, which would be excluded under a component-count threshold. By contrast, a beam-size criterion, selected with guidance from tracers, remains sensitive to extended morphology regardless of the number of fitted components.

During the active learning phase, the tracers are used to train the model in a way that mimics how an expert would explore an unlabelled dataset. By labelling known examples that appear during the labelling iterations, the algorithm learns to prioritise sources with similar diffuse morphologies. This approach allows someone with less experience to label effectively, using the tracers as a reference for what an expert would consider interesting.

The tracers also provide a practical benchmark for assessing performance. By tracking how many appear cumulatively in the top \textit{N} ranked sources, we can see how efficiently the pipeline recovers relevant diffuse emission for inspection. Using this combination of tracer guidance and iterative labelling, \textsc{Protege} can identify rare or subtle sources without relying on classification approaches.

%
%% ====== Subsection
\subsection{Methodological Choices}
\label{sec:methodological choices}
%% ===============
%

The use of PyBDSF for source extraction proved surprisingly effective when islands instead of Gaussian components are used, but has limitations for faint diffuse structures. Default parameters can miss low-SNR sources or fragment very extended emission into multiple components. While fragmentation alone does not prevent detection, as each fragment can still rank highly if it displays diffuse characteristics, it does affect how the structures are represented and can complicate the interpretation of the results. A single radio halo fragmented into three PyBDSF islands might appear as three separate candidates in the ranked list rather than one coherent structure. Alternative source extraction tools or carefully tuned parameters may also improve completeness for faint emission in future applications, although this would require survey-specific optimisation. For example, the \textit{a-trous} wavelet decomposition mode in PyBDSF could enhance the detection of faint, extended emission such as the halos, and investigating its effect on recovery rates is a potential direction for future work.

The pipeline also operates directly on the standard MGCLS enhanced image products without applying prior point-source subtraction. Because of this, when a PyBDSF island is dominated by an embedded compact source, the resulting cutout is heavily influenced by the compact component. In the BYOL feature space, these contaminated cutouts resemble more common compact sources rather than diffuse emission, causing them to receive low scores during active learning. This is illustrated by the candidate diffuse emission in Abell 22 (\autoref{sec:investigation of tracers in lower-ranked positions} and \autoref{fig:abell22_comparison}), where the diffuse emission is faint and ambiguous in the cutout. Even with contaminated sources, the pipeline still recovers a substantial fraction of all of the targeted diffuse emission. We chose this direct approach because it requires minimal manual intervention and preprocessing, making it suitable for very large datasets. Incorporating standard point-source subtraction techniques into the image preprocessing stage prior to source extraction would be a good step to explore to recover the contaminated structures, but it must be approached with caution to avoid removing real emission or introducing artefacts.

We also used a binary labelling scheme, scoring tracers as 5 and all other sources as 0, rather than the 0 to 5 gradient used in previous \textsc{Protege} applications. While this simplifies labelling and reduces user bias, it limits the algorithm because no gradient exists to guide learning from ambiguous cases. For real-world exploration tasks, a gradient-based approach, marking clearly interesting sources as 5, ambiguous candidates as 2 to 3, and uninteresting sources as 0, would provide a stronger and more generalisable learning signal for Gaussian Processes.

%
%% ====== Subsection
\subsection{Feature Space Structure}
\label{sec:feature space structure}
%% ===============
%

The UMAP projections of the high-resolution and convolved feature spaces (see \autoref{fig:UMAP of the high and convolved subsets}) show that tracers form multiple distinct groupings, with some sources dispersed throughout the embedding. This distribution reflects the morphological diversity of the tracer population, which includes both compact or barely resolved sources and highly extended emission.

Because the tracers occupy separate regions in feature space, active learning must explore multiple areas to recover all known sources. Tracers embedded among the general population are particularly challenging, as the algorithm prioritises regions similar to previously labelled examples. This scattered distribution helps explain the presence of lower-ranked sources and  why some of the cumulative plots plateau. Morphologically distinct tracers that appear as outliers in the embedding are less likely to be identified based solely on similarity to previously labelled examples. This effect is particularly evident in the ``180 labels'' iteration of the high-resolution dataset, where cumulative recall temporarily decreases compared to ``120 labels.'' 

This likely reflects how the Gaussian process in \textsc{Protege} samples the feature space: if many labels come from a single cluster, the algorithm may overestimate the importance of that region and reduce sampling elsewhere, temporarily under-representing interesting sources. As further iterations rebalance sampling, the cumulative recall recovers. This emphasises the importance of maintaining a balanced selection of labelled examples during the active learning process.

%
%% ====== Subsection
\subsection{Diffuse Emission versus Cluster-Associated Emission}
\label{sec:detecting diffuse emission vs cluster-associated emission}
%% ===============
%

An important distinction emerges from our results: the algorithm successfully identifies diffuse radio emission in \textit{general}, not exclusively cluster-associated diffuse emission. Of the top 100 ranked sources, 53 are confirmed cluster-related diffuse emission (the tracers), while 45 are other forms of diffuse emission including radio AGN with jets and lobes, bent-tail and wide-angle-tail radio galaxies, face-on spirals with star-forming regions, and unidentified extended structures.

This outcome reflects a fundamental limitation of morphology-based classification using only radio imaging: distinguishing cluster-associated diffuse emission from other extended structures requires additional contextual information beyond what BYOL features can capture. The catalogue by \cite{kolokythas2025} identifies cluster emission by incorporating multiple pieces of evidence including cluster centre positions, X-ray data indicating the presence of hot ICM, spectroscopic redshifts establishing cluster membership, and the spatial correlation between radio emission and cluster properties. Our pipeline, using only radio morphology-based information, does not include this contextual information. Without this contextual information, distinguishing genuine diffuse emission from imaging artefacts, such as radial spikes associated with bright radio sources, can also be challenging, as shown in \autoref{fig:top 100 high res concatenated}.

We therefore view the 45 non-tracer sources not as failures (or false positives) but as successes in identifying candidates worthy of follow-up investigation. Each represents a genuine detection of extended emission that could be of interest. Some may prove to be cluster-associated emission not included in the reference catalogue (either new detections or structures that were ambiguous in the original classification). Others may be interesting for different reasons, such as unusual radio galaxy morphologies, which often indicate environmental interactions or evolutionary stages worthy of study.

This distinction has important implications for interpreting the performance metrics used. The cumulative curves were calculated using the tracers only, reflecting the ability of the algorithm to recover these sources. However, the top 100 ranked candidates include both cluster-associated and non-cluster diffuse sources. As a result, the cumulative recall may underestimate the effectiveness of the algorithm in highlighting promising diffuse emission, as the top-ranked sources may contain genuine cluster emission alongside other interesting diffuse phenomena. Expert validation or multi-wavelength follow-up is therefore required to confirm the nature of these candidates.

What ``successful detection'' means in this study is therefore subjective: \textsc{Protege} excels at prioritising diffuse morphologies broadly even when trained on cluster-related emission only, accelerating the discovery of rare and morphologically diverse sources, but does not perform definitive classification of cluster emission. Future work could integrate complementary information, such as redshifts, cluster membership, multiple radio frequencies for spectral information, or optical and X-ray data, to refine the prioritisation towards genuine cluster-associated emission, as highlighted in \citet{kolokythas2025}. By explicitly acknowledging these limitations, we can better interpret the performance metrics and guide follow-up investigations effectively.

%
%%
%%% ====== Section
\section{Conclusions}
\label{sec:conclusion}
%%% ===============
%%
%

The detection of diffuse emission in galaxy clusters remains a challenge in modern radio astronomy, particularly as data volumes continue to grow with new instruments. Our study demonstrates that combining self-supervised feature learning (BYOL) with active learning (\textsc{Astronomaly: Protege}) provides an effective approach for identifying diffuse radio emission in large astronomical datasets. Applied to the MGCLS, our pipeline substantially reduces inspection effort while recovering the majority of known diffuse cluster emission and identifies additional diffuse candidates.

Three key findings emerge from this work:

First, active learning proves well-suited for rare, morphologically diverse phenomena where traditional supervised approaches struggle due to limited labelled data. By operating iteratively within a single dataset rather than training for generalisation, \textsc{Protege} efficiently adapts to user-defined scientific interests through minimal human feedback (300 labels across 20 iterations).

Second, resolution-dependent features capture complementary information. High-resolution imaging enables the rapid identification of prominent structures, while convolved images enhance sensitivity to faint extended emission. Concatenating features from both resolutions combines these strengths, suggesting that multi-modal approaches incorporating additional wavelengths (X-ray, optical) or radio frequencies could further improve performance.

Third, morphology-based classification using radio imaging alone can not reliably distinguish cluster-associated diffuse emission from other extended structures (radio galaxies or star-forming regions for example). This fundamental limitation highlights the need for incorporating contextual information such as cluster positions, X-ray data, or spectroscopic redshifts, to refine candidate classifications, particularly for large-scale surveys where the manual validation of all candidates becomes impractical.

Several directions warrant future investigation. Gradient-based labelling schemes (0-5 scale) would better capture ambiguous cases than our binary approach, potentially improving sensitivity to morphological edge cases. Transfer learning experiments could test whether BYOL features trained on MGCLS generalise to other surveys or require retraining for different frequencies and resolutions. Hybrid approaches combining our method for initial candidate identification with supervised models for detailed classification could incorporate the strengths of both paradigms.

As radio surveys scale toward SKA data volumes, methods that balance automation with human expertise will become essential. Our demonstration that self-supervised learning combined with targeted active learning can efficiently identify rare phenomena provides a foundation for addressing these challenges. 

Our approach excels for exploratory science targeting rare, poorly-defined morphologies, precisely where supervised methods struggle due to lack of well-defined training data.

%
%%
%%% ====== Section
\section*{Acknowledgements}
\label{sec:acknowledgements}
%%% ===============
%%
%

ML, VE, Kenda Knowles and Konstantinos Kolokythas acknowledge support from the South African Radio Astronomy Observatory and the National Research Foundation (NRF) towards this research. Opinions expressed and conclusions arrived at, are those of the authors and are not necessarily to be attributed to the NRF.
This research made use of data from the MeerKAT telescope, a South African Radio Astronomy Observatory (SARAO) facility. SARAO is a facility of the National Research Foundation, an agency of the Department of Science and Innovation. We acknowledge use of the Inter-University Institute for Data Intensive Astronomy (IDIA) data intensive research cloud for data processing and storage. IDIA is a partnership of the University of Cape Town, the University of Pretoria, and the University of the Western Cape.

ET acknowledges support from the Swiss National Science Foundation under the SNSF Starting Grant ``Deep Waves'' (218396) and travel and collaboration support from the SNSF Weave/Lead Agency Grant ``RadioClusters'' (214815).

This work made use of the following software packages: Astronomaly \citep{Lochner2021}, PyBDSF \citep{2015ascl.soft02007M}, scikit-image \citep{van_der_walt_2014}, scikit-learn \citep{scikit-learn}, umap-learn \citep{UMAP2021_python}, PyTorch \citep{PyTorch}, NumPy \citep{Harris_2020}, Matplotlib \citep{matplotlib}, and Astropy \citep{2013A&A...558A..33A,2018AJ....156..123A}.

%
%%
%%% ====== Section
\section*{Data Availability}
\label{sec:data availability}
%%% ===============
%%
%

The MeerKAT Galaxy Cluster Legacy Survey data used in this study are publicly available through the SARAO archive at \url{https://archive.sarao.ac.za}. The reference catalogue of diffuse radio emission is published in \cite{kolokythas2025}. The Astronomaly framework and Protege extension are open-source and available at \url{https://github.com/MichelleLochner/astronomaly}. 

%
%%
%%% ====== Section
\section*{Conflict of Interest}
\label{sec:conflict of interest}
%%% ===============
%%
%

The authors declare no conflict of interest.

%%%%%%%%%%%%%%%%%%%%%%%%%%%%%%%%%%%%%%%%%%%%%%%%%%

%%%%%%%%%%%%%%%%%%%% REFERENCES %%%%%%%%%%%%%%%%%%

% The best way to enter references is to use BibTeX:

\bibliographystyle{rasti}
\bibliography{bibliography}

%%%%%%%%%%%%%%%%%%%%%%%%%%%%%%%%%%%%%%%%%%%%%%%%%%

%%%%%%%%%%%%%%%%% APPENDICES %%%%%%%%%%%%%%%%%%%%%

\appendix

%
%%
%%% ====== Section
\section{Additional Material}
\label{sec:additional material}
%%% ===============
%%
%

%
%% ====== Subsection
\subsection{Pipeline Flowchart}
\label{sec:appendix flowchart}
%% ===============
%

\begin{figure*}
\centering
    \includegraphics{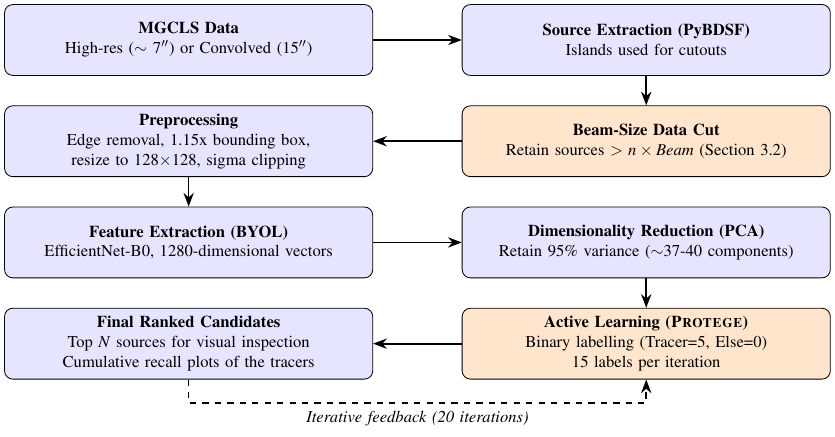}
    \caption{Flowchart of the targeted machine learning pipeline for detecting diffuse radio emission. Steps highlighted in orange indicate the specific methodological adaptations introduced in this work compared to \citet{lochner2024}. During preprocessing, the bounding box values used here also differed compared to \citet{lochner2024}.}
    \label{fig:pipeline_flowchart}
    Alt text: Flowchart of eight boxes joined by arrows, arranged in four rows of two and read in a zigzag, back and forth pattern. The steps are: MGCLS data at high resolution (about 7 arcsec) or convolved (15 arcsec); source extraction with PyBDSF, using islands for cutouts; a beam-size data cut; preprocessing (edge removal, bounding box enlarged 1.15 times, resizing to 128 by 128 pixels, sigma clipping); feature extraction with BYOL, giving 1280-dimensional vectors; PCA keeping 95 per cent of the variance (about 37 to 40 components); active learning with Protege, with binary labels and 15 labels per iteration; and final ranked candidates for visual inspection and recall plots. A dashed arrow from the last step goes back to the active learning step indicating the iterative feedback process with 20 iterations. The beam-size cut and active learning boxes are shaded orange; the rest are blue.
\end{figure*}

\autoref{fig:pipeline_flowchart} illustrates the workflow of the pipeline developed in this work. While the core feature extraction and active learning framework follow \citet{lochner2024} and \citet{Mohale2024}, this diagram highlights the specific adaptations made for the detection of diffuse cluster emission, namely the beam-size based data cut (\autoref{sec:beam cut application}) and the binary labelling scheme (\autoref{sec:active learning with protege}).

%------------------------------------------------%
%------------------------------------------------%

%
%% ====== Subsection
\subsection{Morphology-Specific Recovery Rates}
\label{sec:morphology}
%% ===============
%

The values in \autoref{tab: morphology_comparison} represent the fractional recovery rate, defined as the percentage of known tracers belonging to a specific morphological class, as defined in \citep{kolokythas2025}, that appear within the top $N$ ranked sources output by \textsc{Protege}. For example, a value of 34.4\% for Radio Halos in the ``Mixed / Top 100'' column indicates that 11 of the 32 known halos were ranked in the top 100 positions when the algorithm was trained on all tracer types.

The ``Mixed'' columns represent the standard pipeline presented in this work with active learning applied to all of the known tracers, while the ``Halo-Only'' columns represent a smaller experiment where the active learning algorithm was applied to the tracers corresponding to radio halos only, with all other sources (including other diffuse emission types) treated as uninteresting.

%%%%%%%%%%%%%%%%%%%%%%%%%%%%%%%%%%%%%%%%%%%%%%%%%%
\begin{table*}
    \centering
    \caption{Fractional recovery (\%) of known tracers at different ranking thresholds for the high-resolution concatenated subset. The left side of each block (Mixed) shows the main pipeline with active learning using all of the tracers, while the right side (Halo-Only) shows the results when applying active learning exclusively on the 32 halo tracers.}
    \label{tab: morphology_comparison}
    \begin{tabular}{l cc cc cc cc cc}
        \toprule
        & \multicolumn{2}{c}{\textbf{Top 50}} & \multicolumn{2}{c}{\textbf{Top 100}} & \multicolumn{2}{c}{\textbf{Top 200}} & \multicolumn{2}{c}{\textbf{Top 500}} & \multicolumn{2}{c}{\textbf{Top 1000}} \\
        \cmidrule(lr){2-3} \cmidrule(lr){4-5} \cmidrule(lr){6-7} \cmidrule(lr){8-9} \cmidrule(lr){10-11}
        \textbf{Class (cutouts)} & Mixed & Halo-Only & Mixed & Halo-Only & Mixed & Halo-Only & Mixed & Halo-Only & Mixed & Halo-Only \\
        \midrule
        Radio Halo (32)       & 21.9 & 3.1  & 34.4 & 6.2  & 34.4 & 6.2  & 37.5 & 6.2  & 43.8 & 25.0 \\
        Radio Relic (62)      & 43.5 & 4.8  & 54.8 & 4.8  & 54.8 & 12.9 & 66.1 & 16.1 & 72.6 & 19.4 \\
        Halo + Relic (5)      & 80.0 & 20.0 & 80.0 & 20.0 & 80.0 & 40.0 & 100.0 & 60.0 & 100.0 & 60.0 \\
        Mini-halo (10)        & 10.0 & 0.0  & 10.0 & 0.0  & 20.0 & 0.0  & 30.0 & 10.0 & 30.0 & 20.0 \\
        Phoenix (4)           & 25.0 & 0.0  & 25.0 & 0.0  & 50.0 & 0.0  & 75.0 & 0.0  & 100.0 & 0.0  \\
        Unknown (3)           & 66.7 & 0.0  & 66.7 & 0.0  & 66.7 & 0.0  & 66.7 & 0.0  & 66.7 & 33.3 \\
        \bottomrule
    \end{tabular}
\end{table*}
%%%%%%%%%%%%%%%%%%%%%%%%%%%%%%%%%%%%%%%%%%%%%%%%%%

The ``Mixed'' columns in \autoref{tab: morphology_comparison} show that recovery depends strongly on the location of a source within the BYOL feature space. Visual inspection of the UMAP plot reveals that the relics form more compact groupings at the edges. In contrast, halos and mini-halos are sparsely distributed and overlap heavily with the background population (as discussed in \autoref{sec:discussion} regarding similarity searches), leading to lower recovery rates.

To test whether restricting the active learning labels to a single class could overcome this, we conducted an experiment training exclusively on the 32 known halos (the ``Halo-Only'' columns). Strikingly, halo recovery collapsed from 34.4\% to just 6.2\% within the top 100. We attribute this failure to a combination of factors: (1) the reduction in number of samples reduces the ability of the Gaussian Process to define a boundary; (2) the inherently high overlap of halos with the background sources makes them poor standalone trainers; and (3) halos that were previously recovered in the mixed run resided in regions of the feature space close to relic clusters, and thus lost their high scores when relics were removed from the active learning phase. This confirms that the bottleneck for detecting highly ambiguous morphologies lies in the feature representation itself rather than the active learning labelling strategy.

%
%% ====== Subsection
\subsection{Model Validation and Loss Curves}
\label{sec:model validation and loss curves}
%% ===============
%

%------------------------------------------------%
\begin{figure}
    \begin{minipage}{\columnwidth}
    \centering
    \includegraphics[width=\columnwidth]{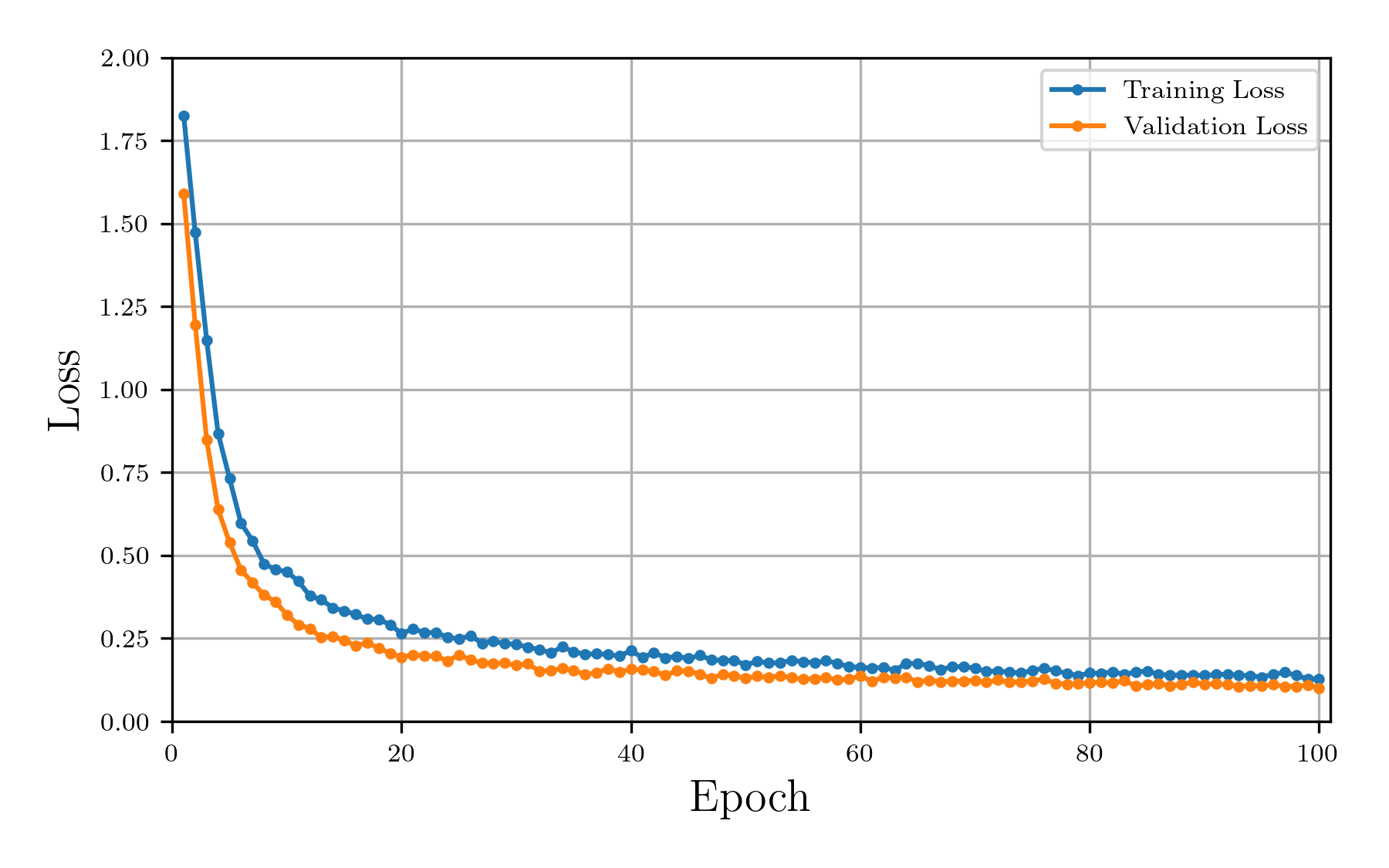}
    \caption{Training and validation loss curves for the BYOL model. The validation set was kept aside to monitor generalisation performance. The alignment of the two curves indicates effective representation learning without overfitting.}
    \label{fig:loss_validation}
    Alt text: Line plot of loss (0 to 2) against training epoch (0 to 100) for the BYOL model, with one training curve and one validation curve. Both start high, at about 1.8 for training and 1.6 for validation, and drop steeply over the first 10 or so epochs. They then decrease slowly and level off at roughly 0.1 to 0.15 by epoch 100. The validation curve stays slightly below the training curve, and the two nearly overlap by the end.
\end{minipage}
\end{figure}
%------------------------------------------------%

During training, a subset of the data was reserved as a validation set for BYOL to monitor generalisation performance. \autoref{fig:loss_validation} shows that both the training and validation loss decrease steadily and converge, confirming that the model learns robust representations from the data without overfitting.

%
%% ====== Subsection
\subsection{Illustration of Algorithm Improvement}
\label{sec:illustration of algorithm improvement}
%% ===============
%

\begin{figure*}
\begin{minipage}{\textwidth} 
    \centering
    \includegraphics[width=0.925\textwidth]{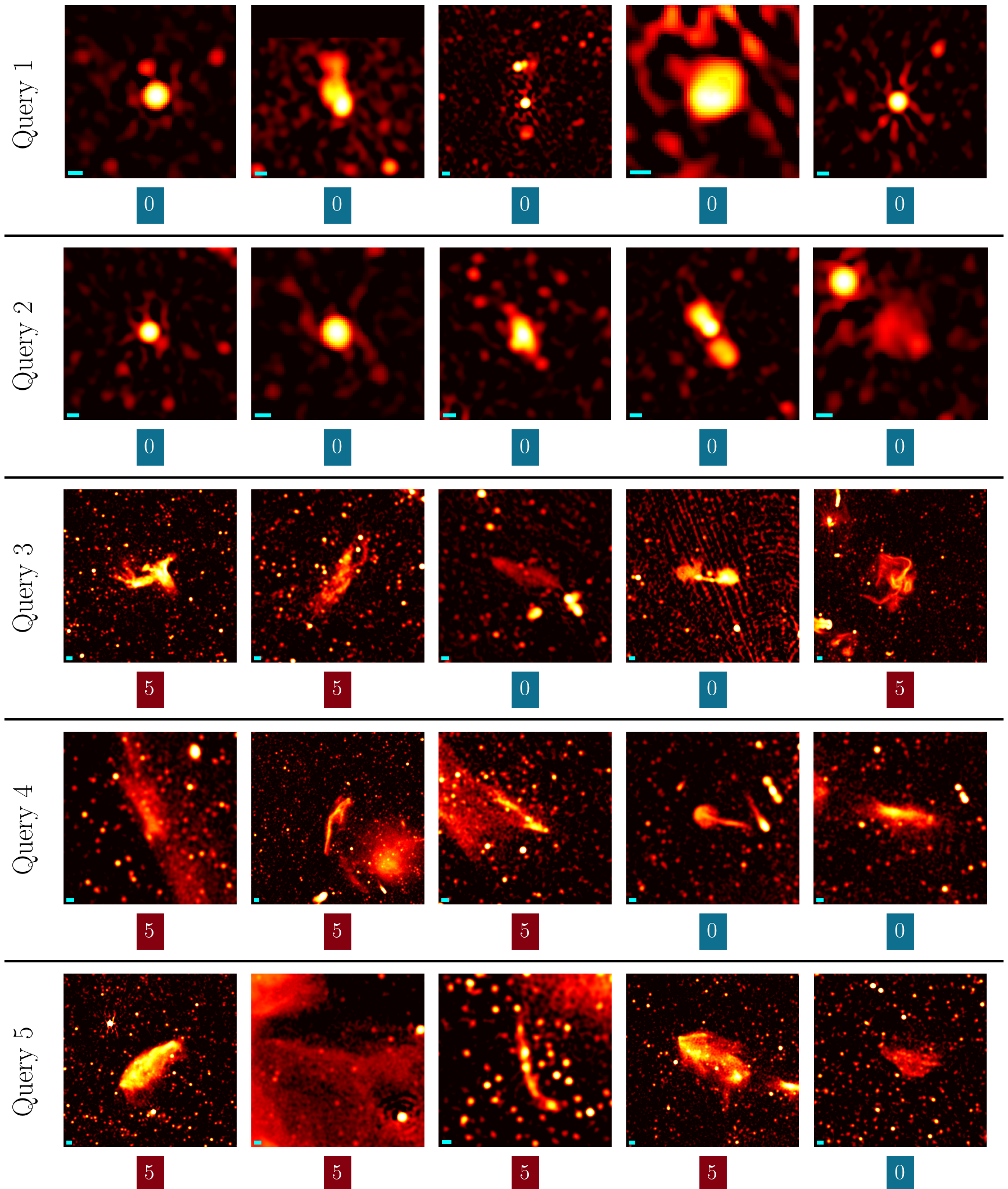}
    \caption{Top 5 sources returned for successive queries during active learning. Each row corresponds to one query iteration, with labels (scores) shown below each source. Over time, more interesting sources are identified and appear earlier in the ranked results, illustrating the effectiveness of the active learning strategy.}
    \label{fig:improvement_with_iteration}
    Alt text: Five rows, labelled Query 1 to Query 5, each with five cutouts and the score given to each (0 or 5) in a box below it. In Queries 1 and 2, the images show mostly compact, point-like sources, and all ten are scored 0. From Query 3 onward, the images show extended structures such as arcs, filaments and diffuse patches, and more are scored 5: three of five in Query 3, three in Query 4 and four in Query 5. A short cyan bar in the lower-left corner of each image marks the beam size.
\end{minipage}
\end{figure*}

\autoref{fig:improvement_with_iteration} illustrates the evolution of the active learning process, analogous to Figure 8 in \citet{lochner2024}. The first row shows the initial random sample, while subsequent rows show the top-ranked sources after each labelling iteration. As the algorithm receives more feedback, it learns to identify and prioritise diffuse emission sources more effectively, as evidenced by the increasing prevalence of interesting sources in the top-ranked positions.

%------------------------------------------------%
%------------------------------------------------%

%
%% ====== Subsection
\subsection{Computational Resources}
\label{sec:compute}
%% ===============
%

All experiments were conducted on a laptop equipped with 32GB of RAM, and an NVIDIA RTX 4060 GPU (8 GB VRAM). Source extraction using PyBDSF required 3 to 20 minutes per field, depending on complexity. Processing was parallelised across 16 threads, reducing the total wall-clock time for the 115 MGCLS fields to approximately 1.2 hours, with a peak memory usage of 6 GB. 

The most computationally demanding step is the model training. For the largest dataset, training was limited by the GPU VRAM, requiring a batch size of 64. Under these constraints, a single epoch took approximately 4.3 minutes, resulting in a total training time of roughly 7 hours for 100 epochs. Peak memory usage was 5.9GB of VRAM. The smaller subsets took significantly less time (20-30s per epoch). Subsequent steps were fast: dimensionality reduction via PCA took less than a minute, and the active learning phase (updating the Gaussian Process and source rankings) required between 0.25s and 15s per iteration, using less than 1.2GB of RAM. 

In practice, the time required for active learning depends on the user, specifically the time taken to visually inspect and classify the sources. A more detailed analysis of human labelling time can be found in \citet{Etsebeth2024}. For this study, the labelling process was automated, removing this variable human component from the total runtime.

\bsp
\label{lastpage}
\end{document}